\documentclass[10pt,journal,compsoc,twoside]{IEEEtran}
\usepackage{booktabs} 
\usepackage{balance}
\usepackage{algorithm,algpseudocode}
\usepackage{amsmath}
\usepackage{graphicx}

\usepackage[caption=false,font=footnotesize]{subfig}
\usepackage{fancyhdr}
\usepackage{mdwlist}
\usepackage{url}
\usepackage{color}
\usepackage{multirow}
\usepackage{amsmath}        

\graphicspath{{incl/}}

\makeatletter
\def\blfootnote{\xdef\@thefnmark{}\@footnotetext}
\makeatother

\usepackage{setspace}
\def\smallerspacecaption{\vspace{-2mm}}

\newcommand{\drop}[1]{}

\ifCLASSOPTIONcompsoc
  \usepackage[nocompress]{cite}
\else
  \usepackage{cite}
\fi

\ifCLASSINFOpdf
\else
\fi

\begin{document}

\newcommand{\maintitle}{A Modern Approach to IP Protection and Trojan Prevention}
\newcommand{\subtitle}{Split Manufacturing for 3D ICs and Obfuscation of Vertical Interconnects}
\newcommand{\thetitle}{\maintitle: \subtitle}

\title{\thetitle}

\author{Satwik~Patnaik,~\IEEEmembership{Student~Member,~IEEE,}
Mohammed~Ashraf,
Ozgur~Sinanoglu,~\IEEEmembership{Senior~Member,~IEEE,}
and~Johann~Knechtel,~\IEEEmembership{Member,~IEEE}

\thanks{A preliminary version of this paper has 
been presented at ICCAD'18~\cite{patnaik18_3D_ICCAD}.}
\IEEEcompsocitemizethanks{\IEEEcompsocthanksitem 
S. Patnaik is with the Department of Electrical and 
Computer Engineering, Tandon School of Engineering, 
New York University, Brooklyn, NY, 11201, USA.\protect\\
\emph{Corresponding authors}: S.\
	Patnaik (sp4012@nyu.edu), O.\ Sinanoglu (ozgursin@nyu.edu), 
    and J.\ Knechtel (johann@nyu.edu).
\IEEEcompsocthanksitem M. Ashraf, O. Sinanoglu, 
and J. Knechtel are with the Division of Engineering, New York University Abu
Dhabi, Abu Dhabi, 129188, UAE.
}}

\markboth{IEEE Transactions on Emerging Topics in Computing}
{Patnaik \MakeLowercase{\textit{et al.}}: \subtitle}

\IEEEtitleabstractindextext{

\begin{abstract}
Split manufacturing (SM) and layout camouflaging (LC) are two promising
techniques
to obscure integrated circuits (ICs) from malicious entities during and after manufacturing.
While both techniques enable protecting the intellectual property (IP) of ICs, SM
can further mitigate
the insertion of hardware Trojans (HTs).
In this paper, we strive for the ``best of both worlds,'' that is we seek to combine the individual strengths of SM and LC.
By jointly extending SM and LC techniques toward 3D integration,
an up-and-coming
paradigm based on stacking and interconnecting of multiple chips, 
we establish a modern approach to hardware security.
	Toward that end,
	we develop a security-driven CAD and manufacturing flow for 3D ICs
in two variations,
one for IP protection and one for HT prevention.
Essential concepts of that flow are (i) ``3D splitting'' of the netlist to
protect,
(ii) obfuscation of the vertical interconnects (i.e., the wiring between stacked chips),
and (iii) for HT prevention, a security-driven synthesis stage.
We conduct comprehensive experiments on DRC-clean layouts of multi-million-gate DARPA and OpenCores designs (and others).
Strengthened by extensive security analysis for both IP protection and HT
prevention,
we argue that entering the third dimension is
eminent for effective and efficient hardware security.
\end{abstract}

}

\maketitle

\renewcommand{\headrulewidth}{0.0pt}
\thispagestyle{fancy}
\lhead{}
\rhead{}
\chead{\copyright~2019 IEEE.
This is the author's version of the work. It is posted here for personal use.
	Not for redistribution.	The definitive Version of Record is published in
IEEE TETC, DOI 10.1109/TETC.2019.2933572}
\cfoot{}

\IEEEdisplaynontitleabstractindextext

\IEEEpeerreviewmaketitle

\IEEEraisesectionheading{\section{Introduction}
\label{sec:introduction}}

\IEEEPARstart{O}{n} the one hand, design practices by the industry attach importance to optimize for power, performance, and area
(PPA) at the level of physical design or
design architecture (e.g., cache hierarchies, speculative execution).
On the other hand, there are powerful attacks such as
\emph{Meltdown}~\cite{lipp18},
which skillfully exploit these very practices to extract sensitive data at runtime.
Besides, malicious foundries may implement so-called hardware Trojans (HTs) which can help an adversary 
to extract sensitive data purposefully~\cite{xiao16}.
Apart from such concerns regarding the security and trustworthiness of hardware at runtime, protecting the hardware
itself from threats such as piracy of design intellectual property (IP) or illegal overproduction is another
challenge.
That is because to avoid the burgeoning cost associated with
ever-shrinking technology nodes,
most chip companies outsource the fabrication of their ICs nowadays to third-party foundries which are potentially untrustworthy.
Moreover, the tools and know-how required for reverse engineering (RE) of even high-end ICs are becoming more accessible and less
costly~\cite{quadir16};
therefore, a malicious end-user obtaining the IP after production is another significant threat.
Various schemes for IP protection have been put forth over the last decade, and most of them can be classified into logic
locking (LL), layout camouflaging (LC), or split manufacturing (SM).
These three classes consider different threats: SM seeks to protect against untrusted foundries, LC against untrusted
end-users, and LL against both.
Accordingly, there are different assumptions on the attackers' capabilities, different
limitations, and different concepts for realization.
We provide more details in Sec.~\ref{sec:background},
and the interested reader may also 
see~\cite{knechtel2019protect}.

Independent of hardware security, 3D integration has made significant
strides
over recent years. The concept of 3D integration is to stack 
and interconnect multiple chips/dies/tiers/layers, thereby promising ``More Moore,'' i.e.,
to overcome the scalability bottleneck which is exacerbated
by ever-increasing challenges for pitch scaling, routing congestion, process variations, et cetera~\cite{
		knechtel17_TSLDM
}.
Recent studies and prototypes have shown that
3D integration 
can indeed offer significant benefits over conventional 2D
chips~\cite{
	fick13,
	kim12_3dmaps}, which can ultimately also help to thoroughly utilize the existing technology nodes.
Besides, 3D integration advances manufacturing capabilities by various means such as parallel handling of wafers, higher yields for the smaller
outlines of individual chips, and heterogeneous integration (``More than Moore'').

\begin{figure}[tb]
\centering
\includegraphics[width=.8\columnwidth]{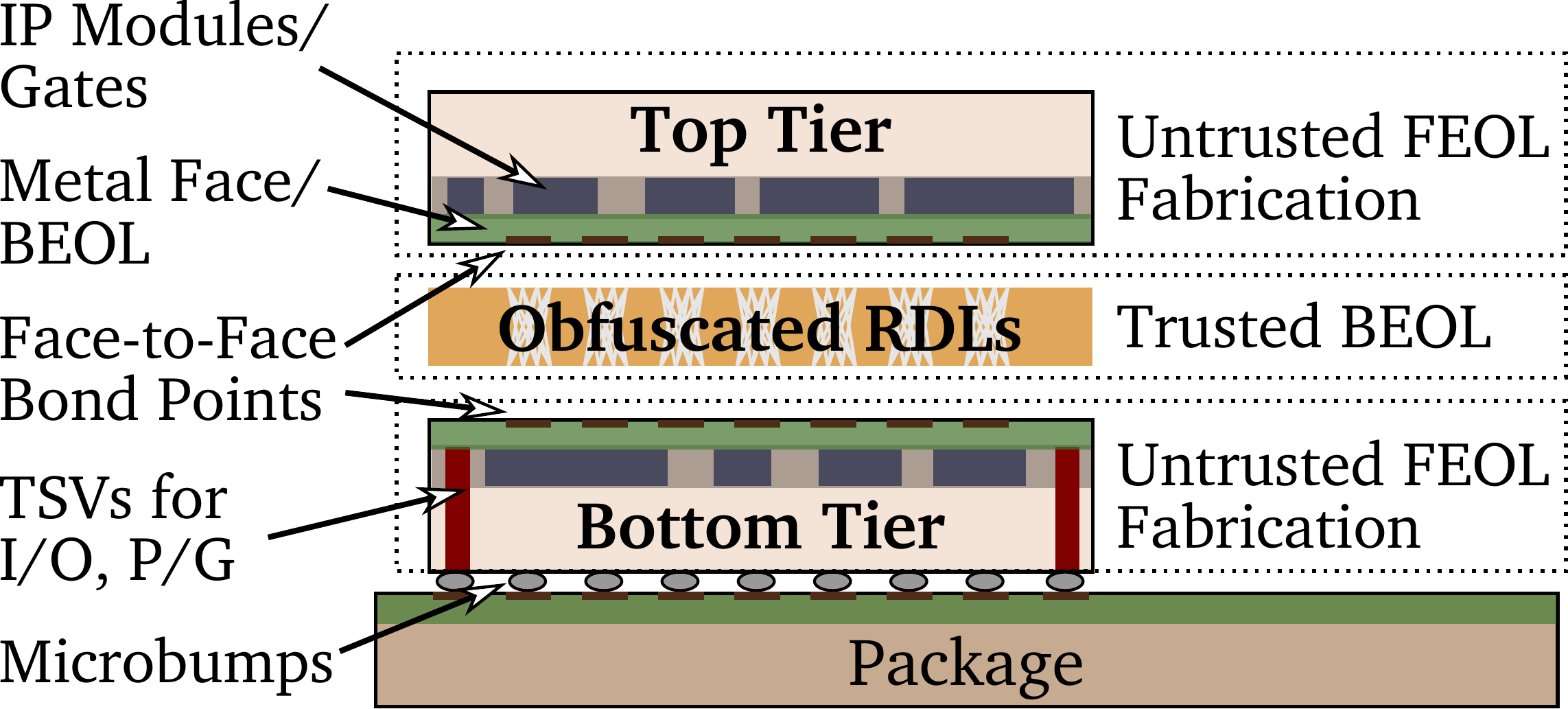}
\caption{Our scheme is based on 3D integration, 
particularly on face-to-face (F2F) 3D ICs.
Through-silicon vias (TSVs) are used for external connections, 
and redistribution layers (RDLs) for vertical interconnects.
Our security-centric splitting of the design across two tiers, along with obfuscation of the RDLs, allows for IP
protection and prevention of HT insertion. That is because the two tiers, even when considered together by colluding foundries, reveal neither
the entire design nor individual components prone to HT insertion.
\label{fig:scheme}
}
\smallerspacecaption
\end{figure}

In this paper, we propose a modern approach to hardware security.
We show that 3D integration is an excellent candidate
to combine the strengths of
LC and SM
in one scheme
(Fig.~\ref{fig:scheme}).
The key idea is to ``3D split'' the design into multiple tiers and to obfuscate (i.e., randomize and camouflage) the vertical interconnects between those
tiers.
Our approach is a significant advancement over prior art in IP protection---while LC may thwart end-user adversaries and SM may safeguard 
against fab-based adversaries, only our work
can readily protect against both threats.
We note that LL targets for the same, but the
viability of LL schemes depends on
tamper-proof memories,
which is an area of ongoing
research~\cite{
	jiang18}.
Concerning HT prevention, the second major part of our work besides IP protection, we show that 3D integration advances the
	state-of-the-art in two ways.
	For one, components considered prone to HT insertion can be delegated to trusted facilities for fabrication of separate
	chips, hindering HT insertion to begin
		with. For another, when the decision on which components are prone to HT insertion is difficult,
		our notion of obfuscating vertical interconnects
			is essential to
		implement other, foundationally secure schemes with superior cost and scalability.

The contributions of this work are as follows:

\begin{enumerate}

\item We put forward a practical threat model which is in line with the present-day business practices of design houses
(Sec.~\ref{sec:practical_threat_model}).
This model necessitates the application of both LC and SM in conjunction.

\item 
We leverage 3D integration for a modern approach to IP protection and HT prevention (Sec.~\ref{sec:concept}).
We combine the strengths of LC, SM, and prior art on HT prevention, all within one concept.
The key idea is to ``3D split'' the design into two tiers (or more, in principle) and to obfuscate the vertical interconnects between those tiers.
We explore two scenarios for commissioning different trusted and untrusted foundries, offering different or same
technology node(s).
In that section and throughout the paper,
	we use \emph{DARPA} and \emph{OpenCores} multi-million-gate
designs
	(besides the well-known \emph{ITC-99} and \emph{ISCAS-85} benchmarks), to 
provide meaningful experimental studies.

\item We develop a security-driven CAD and manufacturing flow for face-to-face (F2F) 3D ICs, initially tailored for IP protection
(Sec.~\ref{sec:IP_methodology}).
In addition to various steps required for an end-to-end 3D IC CAD flow, key concepts for the flow are security-driven partitioning techniques as
well as obfuscation of the vertical interconnects in the F2F 3D IC.
We implement our flow using \emph{Cadence Innovus} and demonstrate its applicability on a broad set of benchmarks.

\item We conduct a thorough analysis of DRC-clean layouts,
    and we contrast with 
the prior art of LC or SM
wherever applicable (Sec.~\ref{sec:IP_results}).
We further present an extensive security analysis, underpinned by analytical and empirical data as well as by a novel proximity-centric attack on 3D ICs
(Sec.~\ref{sec:IP_results}).

\item We extend our flow toward HT prevention (Sec.~\ref{sec:HT_methodology}), specifically for a strong threat model where the fab-based
attacker already holds the full netlist to begin with~\cite{imeson13}.
Initially we identify
the limitations of prior art~\cite{imeson13,li18}: the attainable level of security, scalability, and 
layout cost incurred to guarantee this level of protection.
To tackle these limitations,
we propose and implement a
security-driven synthesis strategy along with our established 3D IC CAD flow.
Here we also enable,
for the first time, the concerned designer to purposefully protect structures of choice at design time.

\item We comprehensively study various scenarios to demonstrate the resilience and efficacy of our approach to HT prevention
(Sec.~\ref{sec:HT_results}).
For example, for the \emph{ITC-99} benchmark b19, 
the attacker has only a chance of 0.25\% 
for successful targeted HT insertion.

\end{enumerate}

\section{Background}
\label{sec:background}

\subsection{3D Integration and CAD Flows}
\label{sec:limitations_3D_CAD}

\begin{figure}[tb]
\centering
\includegraphics[width=\columnwidth]{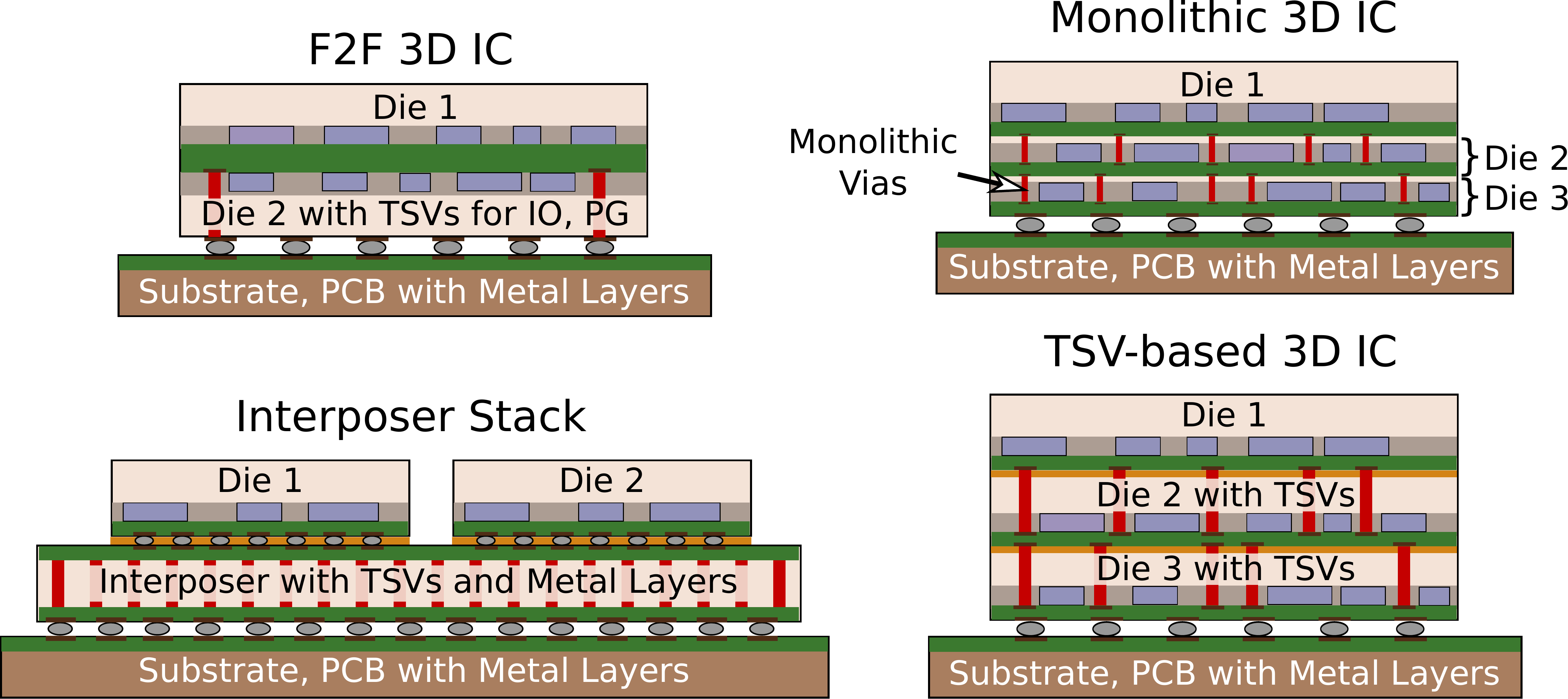}
\caption{The four main flavors of 3D integration. Metal layers are colored in green, active layers in brown (along with modules in blue),
	and bonding underfills in yellow. F2F 3D ICs allow for direct metal-to-metal bonding.
\label{fig:3D_schemes}
}
\end{figure}

3D integration has experienced significant traction over the recent years, for both improving scalability as well manufacturing and
integration capabilities~\cite{
		ku18,chang16,
	knechtel17_TSLDM}.
3D integration can be broadly classified into four flavors (Fig.~\ref{fig:3D_schemes}):
(1) through-silicon via (TSV)-based 3D ICs, where chips are fabricated separately and then stacked, with the vertical inter-chip connections being
realized by relatively large metal TSVs running through the entire silicon chips;
(2) F2F stacking, where two chips or tiers are fabricated separately and then bonded together directly at their metal faces (along with TSVs
		only required for external connections);
(3) monolithic 3D ICs, where multiple tiers
are manufactured sequentially, with the vertical interconnects based on regular metal vias;
(4) 2.5D integration, where chips are fabricated separately and then bonded to a system-level interconnect
carrier, the interposer. Each flavor has its scope, benefits and drawbacks, and requirements for CAD and manufacturing
processes~\cite{
		knechtel17_TSLDM}.

F2F stacking has arguably emerged as the most promising (along with monolithic 3D ICs);
various studies are actively streamlining efforts for commercial adoption, e.g., \cite{ku18,
	peng17,
	fick13}.
Note that prior art is inherently oblivious to hardware security.
These studies
carefully trade off intra-tier wiring with vertical interconnects across tiers.
While vertical interconnects are the key feature of 3D integration, an overly large number of crossings/cuts can have a significant,
counter-productive impact on PPA~\cite{ku18}. As we explain in Sec.~\ref{sec:IP_security}, however, a large number of cuts is
mandatory for a strong resilience against IP piracy. Hence, our flow comprises techniques to explore this security-cost trade-off.

\subsection{Split Manufacturing}
\label{sec:split_manufacturing}

SM offers an interesting solution to safeguard the design IP,
   but
\emph{only during manufacturing time}.
That is, SM cannot protect against malicious end-users.
   Traditionally, SM means that
the device layer and few lower metal layers (front-end-of-line, FEOL) are fabricated using a high-end, 
   potentially \emph{untrusted} foundry, whereas the remaining interconnects (back-end-of-line, BEOL) are grown
   on top of the FEOL wafer by a \emph{trusted} facility.
   Considering the different pitches of the FEOL/BEOL metal layers,
SM supports a cost-aware supply chain.
That is, as long as the FEOL and BEOL are separated at some higher metal layer, any low-end facility may be commissioned for the
BEOL with relatively little commercial cost.\footnote{This consideration is also essential for adopting SM as a security scheme in practice.
It seems more realistic to find some well-established and trustworthy facility which can, however, only offer large-pitch processing capabilities
than to find some high-end facility which can also be trusted, among the few to begin with. See also Sec.~\ref{sec:diff_found}.}
   Now, the security promise of SM lies in the fact that an
   \emph{untrusted} foundry only holds a part of the overall design, making it difficult to infer the complete design functionality, 
   and thereby hindering an adversary from IP piracy or 
   targeted insertion of HTs.

Existing CAD tools, however, due to their focus on design closure 
(and their so-far agnostic view on security), tend to leave 
hints for an FEOL-based adversary. 
For example, to honor PPA, any to-be-connected cells are typically placed close to each other.
Based on this insight,
Rajendran~\emph{et al.}~\cite{rajendran13_split} proposed a so-called proximity attack which models
this principle to infer the missing BEOL connections.

Various placement-centric~\cite{wang18_SM,
	patnaik18_SM_DAC} and/or
routing-centric~\cite{
patnaik18_SM_ASPDAC,patnaik18_SM_DAC} 
schemes have been proposed recently, which all aim to counter the efforts of
proximity attacks~\cite{wang18_SM,
	rajendran13_split}.
Among those defense schemes, lifting of wires above the split layer remains an intuitive way to obfuscate the IP.  That is, the 
critical wires (as selected by the designer) are lifted, e.g., with the help of constraining the router via routing pins in higher layers or
inserting artificial routing blockages.  We conducted exploratory experiments on the randomized lifting of nets
(Fig.~\ref{fig:naive_lifting_overheads}); here we observe steady and significant increases in PPA cost.
   More comparative
results, also on 2.5D/3D solutions, are given in Sec.~\ref{sec:IP_results}.

We acknowledge that the basic idea for 3D SM was already envisioned in 2008 by Tezzaron~\cite{tezzaron08}.
Also, various studies are hinting at 3D integration for SM, but most have limitations or cover different scenarios. For example, Dofe
\emph{et al.}~\cite{dofe17}
remain on the conceptional level, or
Xie \emph{et al.}~\cite{xie17} and
Imeson \emph{et al.}~\cite{imeson13} consider 2.5D integration where only wires are hidden from the untrusted foundry.
We summarize the prior art on 3D SM in Table~\ref{tab:prior_3D_security}, along with that for 3D LC.

\begin{table}[tb]
\centering
\scriptsize
\caption{Comparison of 3D schemes for hardware security.
The integration style is either 2.5D, stacked 3D (3D), or monolithic 3D (M3D).}
\label{tab:prior_3D_security}
\smallerspacecaption
\setlength{\tabcolsep}{0.8mm}
\begin{tabular}{|c|c|c|c|c|c|}
\hline
\textbf{Reference} & 
\textbf{Style} & 
\textbf{Scheme} & 
\textbf{Scope} & 
\textbf{Assets} &
\textbf{Trusted Entity} \\ 
\hline 
\hline
~\cite{imeson13} & 2.5D & SM & HT & Wires & BEOL \\ \hline
~\cite{xie17} & 2.5D & SM & IP Piracy & Wires & BEOL \\ \hline
~\cite{gu2018cost} & 3D & SM \& LC & IP Piracy & Gates \& Wires & FEOL \& BEOL \\ \hline
~\cite{yan17_camo} & M3D & LC & IP Piracy & Gates & FEOL \\ \hline
\bf{Ours} & \bf{3D} & \bf{SM \& LC} & \bf{IP Piracy \& HT} & \bf{Gates \& Wires} & \bf{BEOL} \\ \hline
\end{tabular}
\end{table}

\subsection{Layout Camouflaging}
\label{sec:layout_camo}

LC foils an adversary's efforts for RE of a chip.
LC is accomplished during manufacturing by (i)~dissolving optically distinguishable traits of standard cells, 
e.g., using look-alike gates~\cite{rajendran13_camouflage} or secretly configured MUXes~\cite{wang16_MUX}, (ii)~selective doping implantation for
threshold-voltage-based obfuscation~\cite{
nirmala16,akkaya18}, or (iii)~rendering the BEOL wires and/or vias resilient against RE~\cite{patnaik17_Camo_BEOL_ICCAD}.
It is important to note that most schemes require alterations to the FEOL process, which can be complex and costly.
In any case, since physical obfuscation constitutes the secret for IP protection by LC, the involved manufacturing
facilities have to be \emph{trusted}---LC cannot protect against malicious fabs.

Powerful Boolean satisfiability (SAT)-based attacks~\cite{subramanyan15
}
have questioned the efficacy of various LC schemes,
leading to a ``cat-and-mouse game'' between adversaries and defenders.
The base for these attacks are analytical models for all possible assignments of the camouflaged design parts
and efficient pruning of 
the search space of assignments.
Recent security schemes thus attempt to impose excessively complex problem instances for SAT solvers by
(i) inserting dedicated challenging structures like camouflaged AND trees~\cite{li16_camouflaging}, (ii) minimally modifying critical
parts of the design functionality~\cite{yasin16_CamoPerturb}, or (iii) full-chip camouflaging~\cite{patnaik17_Camo_BEOL_ICCAD}.

\begin{figure}[tb]
\captionsetup[subfigure]{labelformat=empty}
\centering
\subfloat[]{
\includegraphics[width=.47\columnwidth]{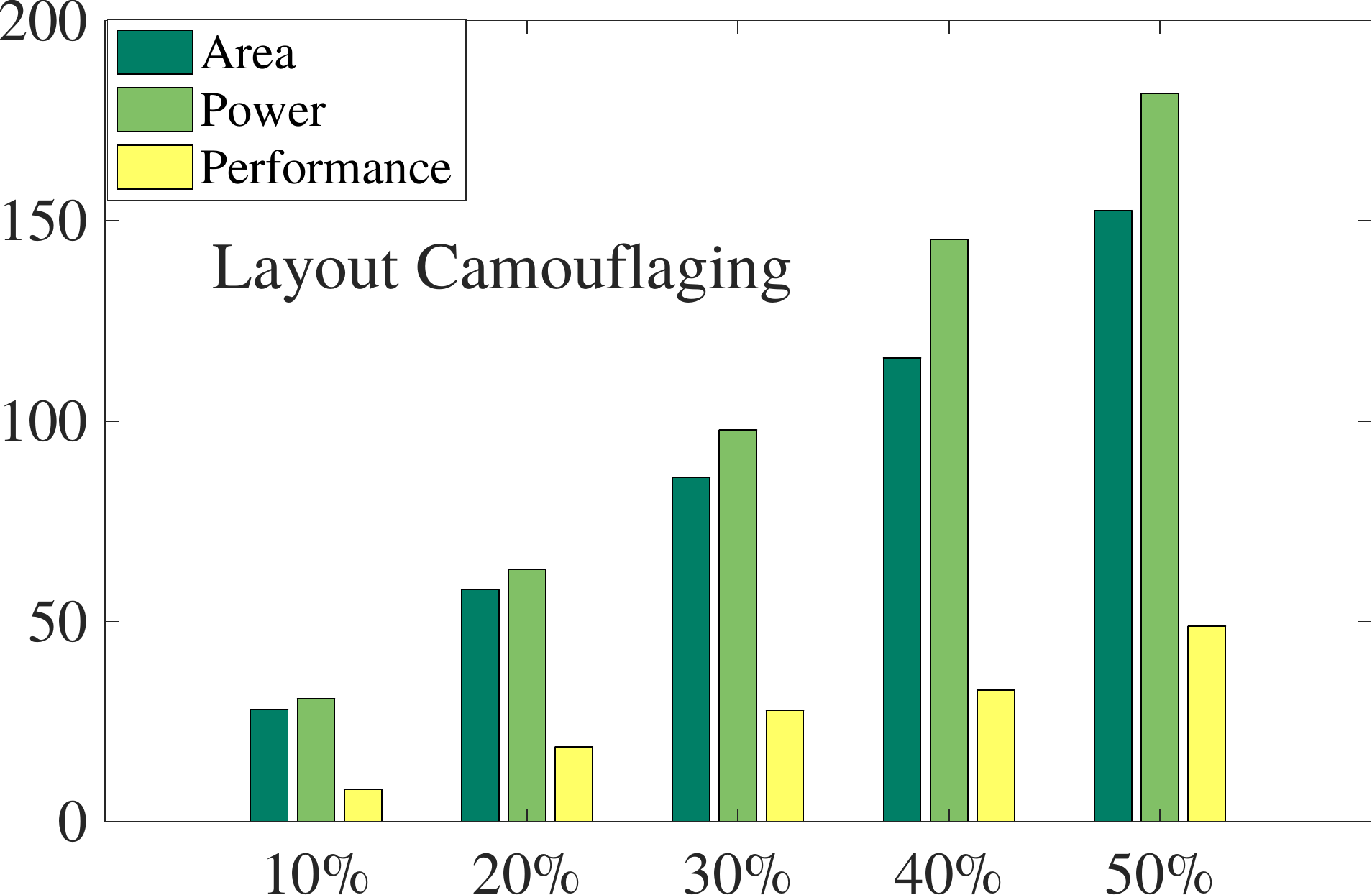}
}
\hfill
\subfloat[]{
\includegraphics[width=.47\columnwidth]{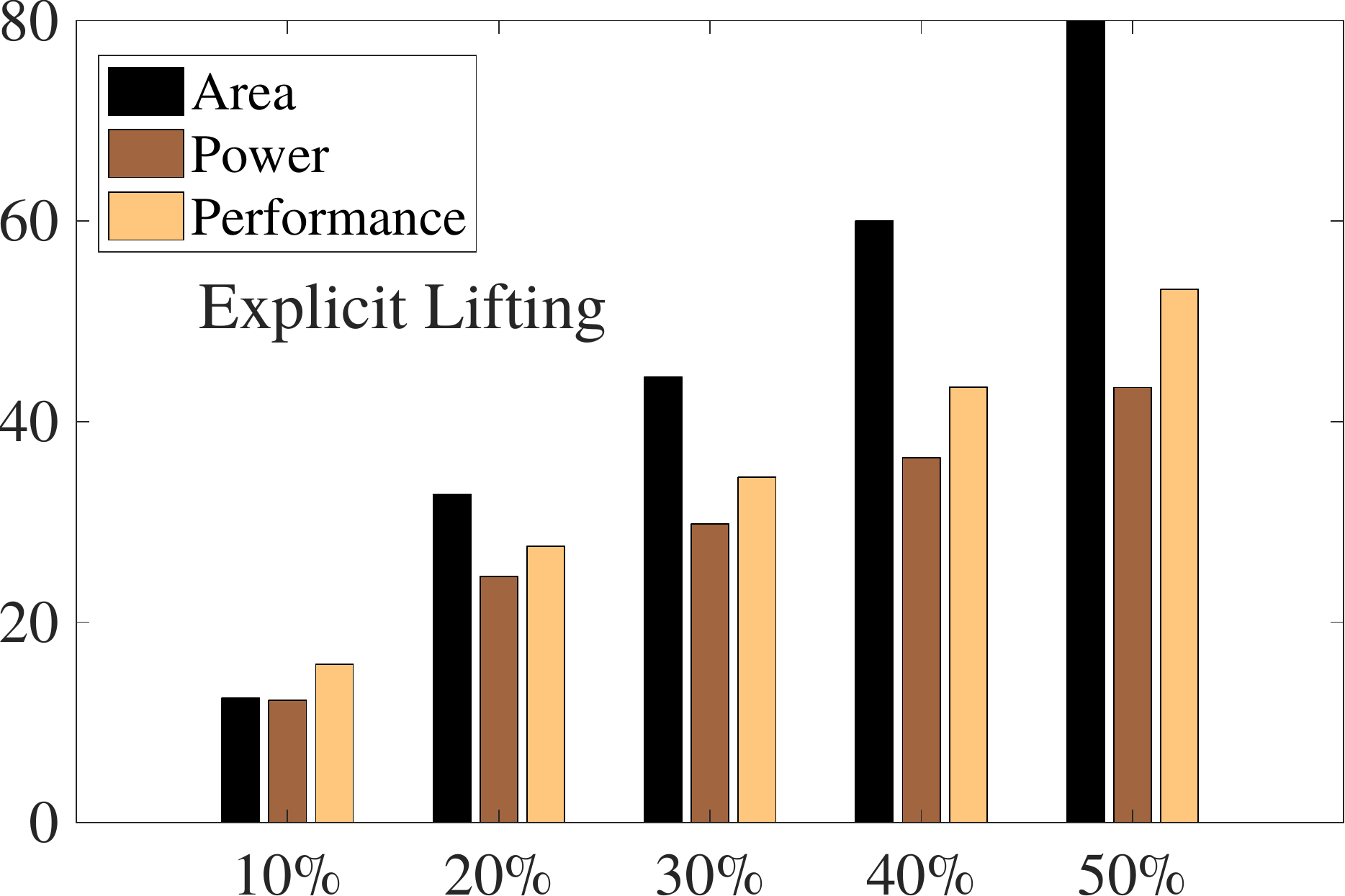}
}
\smallerspacecaption
\caption{PPA cost (\%) for look-alike LC~\cite{rajendran13_camouflage} (left) and explicit lifting of randomly selected wires to
M8 (right) in SM. Results are averaged across
\emph{ITC-99} benchmarks. For the LC scheme (left),
the impact on power and area is substantial, given that the NAND-NOR-XOR structure in~\cite{rajendran13_camouflage} incurs 4$\times$ and
5.5$\times$ more area and power compared to a regular 2-input NAND gate.
For the SM scheme (right), the cost for area is severe. 
That is because routing resources are relatively scarce for M8 (pitch = 0.84$\mu m$) and 
lifting of wires occupies further resources, which can only be obtained by enlarging the die outlines.
\label{fig:naive_lifting_overheads}
\label{fig:LC_overheads}
}
\end{figure}

Existing schemes tend to incur significant PPA overheads once LC is 
applied for large parts of the design.
For example in~\cite{rajendran13_camouflage},
camouflaging 50\% of the design results in $\approx$150\% 
overheads for power and area, respectively (Fig.~\ref{fig:LC_overheads}).
Emerging schemes such as threshold-voltage-dependent LC
still suffer from PPA overheads; see Sec.~\ref{sec:IP_results} for more comparative results.
As for 3D integration, Yan \emph{et al.}~\cite{yan17_camo} proposed LC for monolithic 3D ICs,
and Gu \emph{et al.}~\cite{gu2018cost} apply LC for 3D ICs, albeit using regular 2D LC schemes.
Hence, while promising, both works still require trusted FEOL facilities.

\subsection{Hardware Trojans}
\label{sec:trojans_and_countermeasures}

Apart from IP protection, the possibility for HT insertion by an untrustworthy foundry also raises concerns, especially for
military applications.
ICs that are ``bugged'' with HTs may (i) deviate from their specified functionality, (ii) leak sensitive information, and/or (iii) become
unreliable or fail at particular points in time~\cite{xiao16}.
Trojans can be broadly classified into digital Trojans and physical Trojans,
depending on their payload and
trigger mechanisms.
Digital HTs are activated by
either a specific, rare input pattern or via ``time bombs'' on certain operations (or input patterns) being executed for a particular number of
cycles.  Physical Trojans are activated either by (i) aging effects such as electro-migration, or (ii) internal or external side-channel
triggers.
In this paper, we limit the scope to digital Trojans.

   In general,
there are two classes of HT countermeasures: \emph{reactive}, i.e., monitoring for HTs at runtime,
and \emph{pro-active}, i.e., seeking to
prevent HT insertion during the design and/or manufacturing time.
For reactive countermeasures, there are various approaches, e.g., based on current monitoring~\cite{guimaraes17}
	or
	security wrappers~\cite{basak17}.
For pro-active countermeasures, prior art covers, e.g., built-in self-authentication modules~\cite{shi17_OBISA} and SM-centric
obfuscation schemes~\cite{imeson13,li18}
	among others.
Note that the fab-based adversary needs to comprehend the layout under attack
for targeted HT insertion; that is why schemes like LL and SM can prevent HT insertion to a certain extent.

In this work, we focus on pro-active prevention of HTs.
In particular, we consider a \emph{strong threat model} where the attacker already holds the complete gate-level netlist~\cite{imeson13}.
We review the prior art~\cite{imeson13,li18} in more detail in Sec.~\ref{sec:strong_model_prior_art}.

\section{A Practical Threat Model}
\label{sec:practical_threat_model}

Here we put forward a novel, practical threat model which is in line with the business practices of present-day electronics companies.
Consider the following scenario.
Lacking its own fabrication facilities,
a company commissions a potentially malicious foundry to manufacture their newest version of some chip.
This new version is typically extended
from some previous version (Fig.~\ref{fig:foundry_options}(a))---the reuse of IP modules and the re-purposing of proven
architectures are well-known practices.
For example, think of the flagship iPhone\textsuperscript{\textregistered} by Apple\textsuperscript{\textregistered}. The iPhone
7,
	based on the A10 chip,
	was launched in September 2016, and the
iPhone X, based on the successor chip A11,
was launched in September 2017.
In such a scenario, it is intuitive that pirating the new IP can become significantly less challenging for fab-based
adversaries. In case the same fab was already commissioned for the previous chip version, they readily hold the layout of that earlier version;
otherwise, the adversaries can apply RE on chips of the previous version bought in the market.
In any case, the adversaries can compare that new layout with the prior layout,
to locate and focus on those parts which are different and unique.
Recall that understanding the layout and its functionality in full is necessary for targeted HT insertion; this becomes notably less challenging
as well.

Now, the conclusion for this threat model is that \emph{both LC and SM are required for manufacturing of all different chip versions}.
LC is required to prevent RE of the current layout by any other fab commissioned for later chip
versions, whereas SM is necessary to prevent the fab which is manufacturing the current version (and
which is also tasked to implement LC) from readily inferring the complete layout of the current version.
Prior art can only account for this scenario by applying SM \emph{on top of} LC, which can exacerbate
the individual overheads and shortcomings, as discussed in Sec.~\ref{sec:background} and \ref{sec:IP_results}.
Next, we outline our scheme to combine SM and LC naturally while leveraging 3D integration.

\section{3D Integration as Modern Approach to IP Protection and Trojan Prevention}
\label{sec:concept}

\begin{figure}[tb]
\centering
\includegraphics[width=.85\columnwidth]{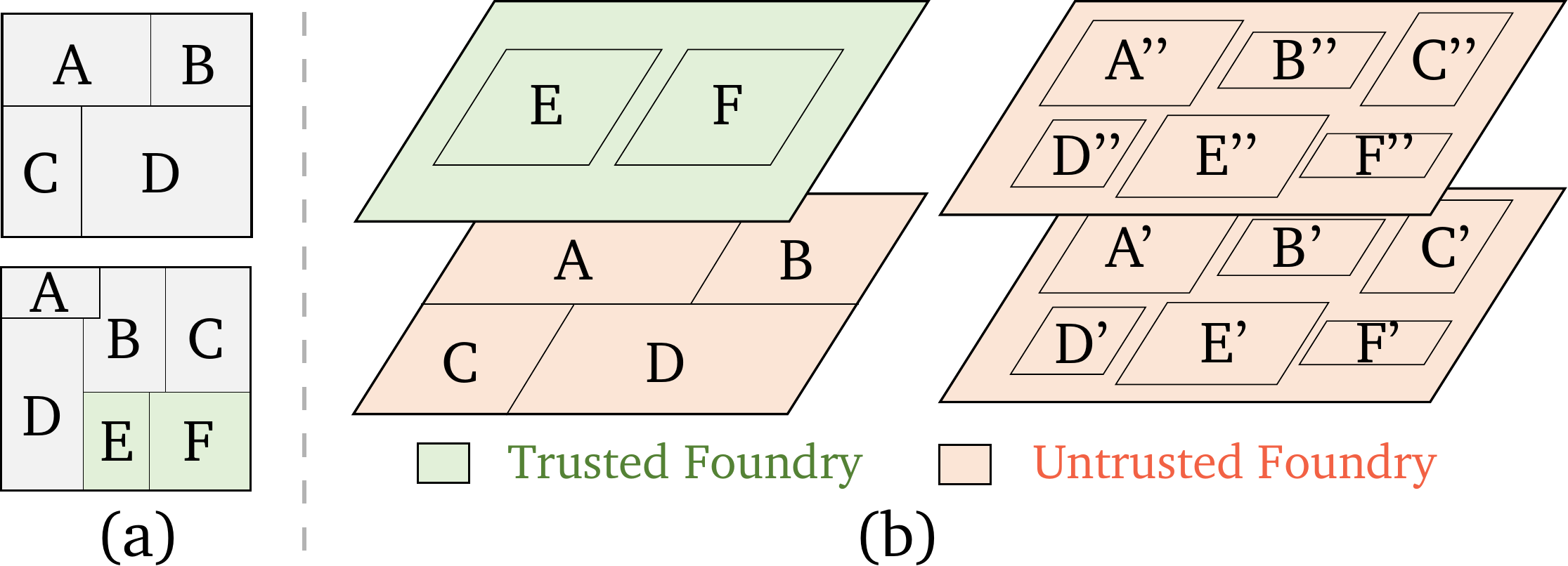}
\caption{(a) Current chip version (top) versus new chip version (bottom).
For the new version, the IP modules \emph{E} and \emph{F} are entirely new, while the other modules are revised and/or reshaped.
(b) Foundry scenarios for our IP protection scheme on 3D ICs.
For both tiers manufactured by an untrusted foundry (right), IP modules can be split up for obfuscation.
\label{fig:foundry_options}
}
\end{figure}

The primary advancement we advocate for SM is to ``3D split'' the design into multiple tiers. That is, unlike regular SM in 2D where the
layout is split into FEOL and BEOL, we split the design itself into two parts (or more, in principle).
These parts are manufactured as separate
chips, and then stacked and vertically interconnected following the F2F integration process (the latter without loss of generality).
We suggest that 3D SM can be achieved either by
commissioning different foundries or one foundry
(Fig.~\ref{fig:foundry_options}(b)):

\begin{enumerate}

\item \emph{Different trusted and untrusted foundries (Sec.~\ref{sec:diff_found}):}
Consider one trusted and one untrusted foundry, both with full FEOL and BEOL capabilities, but for different technology nodes.
It is intuitive to 
delegate the sensitive parts to the trusted fab exclusively.
While this approach
is straightforward and inherently
secure against fab-based adversaries, its practicality is limited, as we discuss below.
\item \emph{Untrusted foundries/foundry (Sec.~\ref{sec:untrusted_foundry}):} Consider one or more high-end but untrusted fab(s). 
This way, we can benefit from the
latest technology but, naturally, have to obfuscate the design in such a way that the fab(s) cannot readily infer the whole layout, even
when they are colluding.
Once such strong protection is in place,
it is economically more reasonable to commission only one fab.
\end{enumerate}

It is important to note the following. First, we elaborate on both scenarios in this section, but we
focus on the more relevant and practical Scenario 2) in the remainder of this work.
Second, to further achieve security against (a) fab-based adversaries and 
(b) malicious end-users, we later on
(a) randomize the vertical interconnects and (b) obfuscate those interconnects. Therefore, we then require a trusted BEOL
facility, but no trusted FEOL facility.

\subsection{Different Trusted and Untrusted Foundries}
\label{sec:diff_found}
\label{sec:naive_3D}

Commissioning several foundries providing different trust levels and supporting different technology nodes
holds two key implications as follows.

\begin{figure}[tb]
\centering
\includegraphics[width=1\columnwidth]{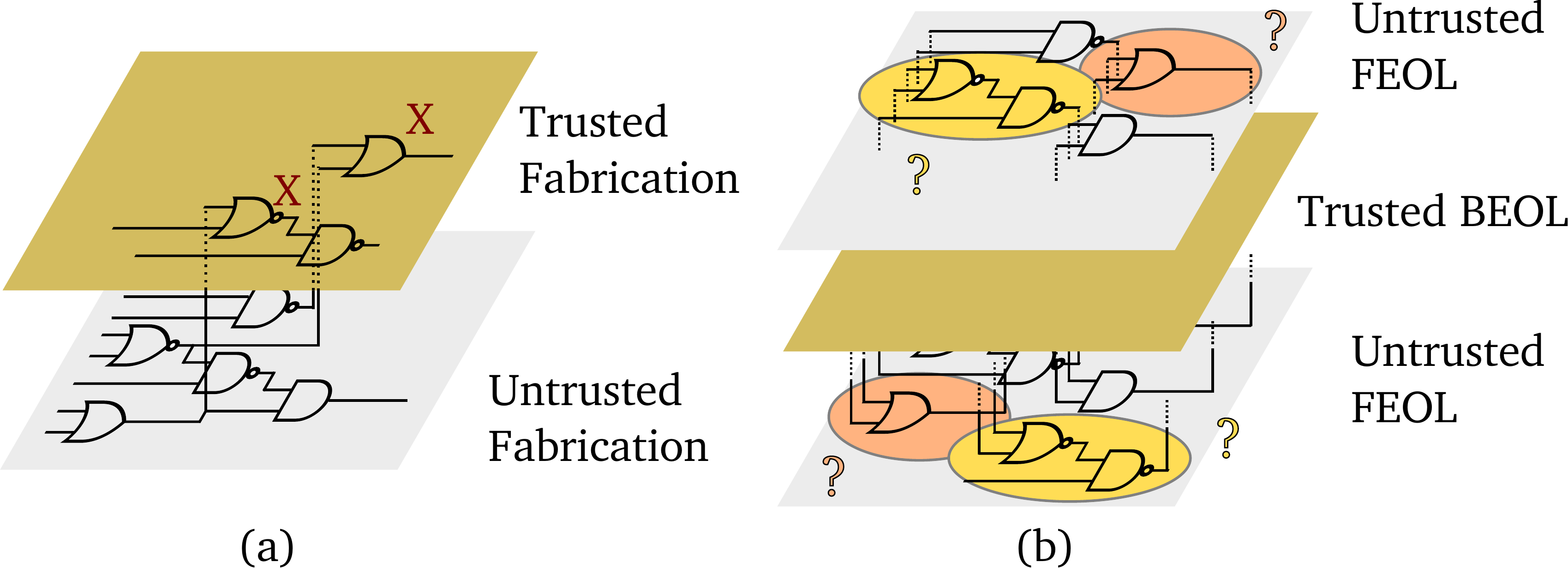}
\smallerspacecaption
\smallerspacecaption
\caption{Two approaches to 3D integration to prevent targeted HT insertion.
(a)~Assuming a trusted foundry (along with another untrusted one),
the vulnerable design parts (marked by ``X'') can be 
delegated to the trusted foundry.
(b) Assuming only an untrusted FEOL facility, the design has to be split in such a manner that an FEOL-based
attacker cannot readily identify the vulnerable parts.
This also requires randomization of the vertical interconnects using a trusted BEOL facility.
\label{fig:HT_3D}
}
\end{figure}

First, it is intuitive to assign sensitive design parts to the chip manufactured by the trusted foundry, for example (i) some new IP to
protect (Fig.~\ref{fig:foundry_options}(b)), or (ii) parts considered vulnerable to targeted HT insertion (Fig.~\ref{fig:HT_3D}(a))---such an
approach is \emph{secure by construction} against fab-based adversaries for the following reasons.
For IP piracy, there is no
generic attack model in the literature yet which can
infer missing connections \emph{and} gates when given only a part of the overall design.
We believe that such ``black-box attacks'' would be very challenging, if possible at all.
For HT insertion,
the adversary cannot perform targeted insertion once the vulnerable parts are delegated to the trusted chip exclusively.

Second, in case a trusted fab and another untrusted fab are commissioned in parallel, it is implied that these two fabs would
support different nodes, with the trusted fab typically offering only access to an older technology.
In fact,
if the trusted foundry would be able to offer the same high-end node, one could simply commission the trusted foundry for
manufacturing of the whole design.
Due to the different pitches for different technology nodes, however, only a fraction of the design can be delegated to the trusted low-end
fabrication.
Also, power and performance will be dominated by the low-end chip, where factors such as parasitics, level shifting, and clock
synchronization may further exacerbate the overheads~\cite{peng17,garg13}.

\subsubsection{Case Study on DARPA Common Evaluation Platform}
\label{sec:CEP_experiments}

\begin{figure}[tb]
\centering
\subfloat[]{\includegraphics[width=.5\columnwidth, trim = {16mm 0mm 0mm 0mm}, clip=true]{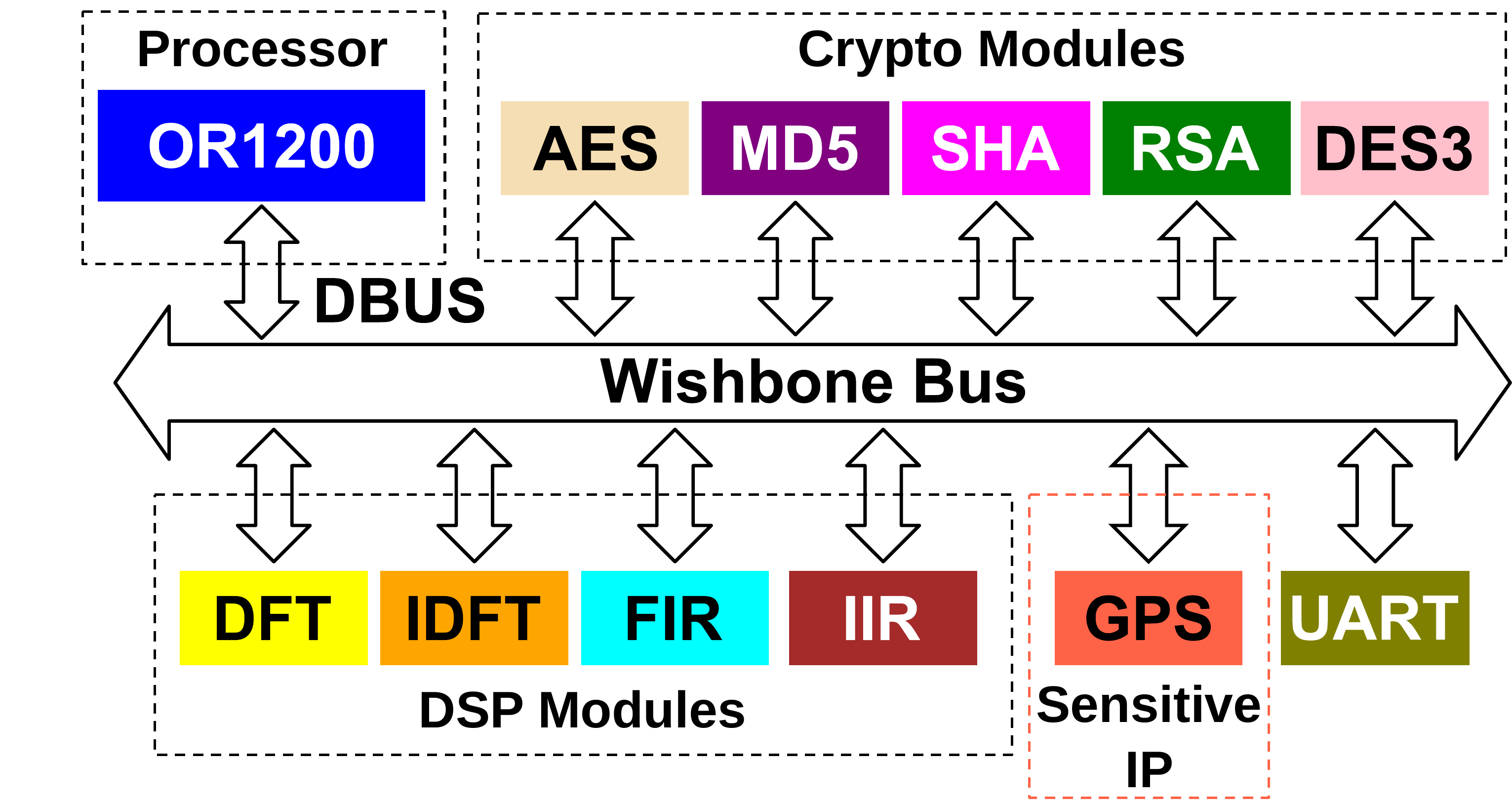}}
\hfill
\subfloat[]{\includegraphics[width=.49\columnwidth]{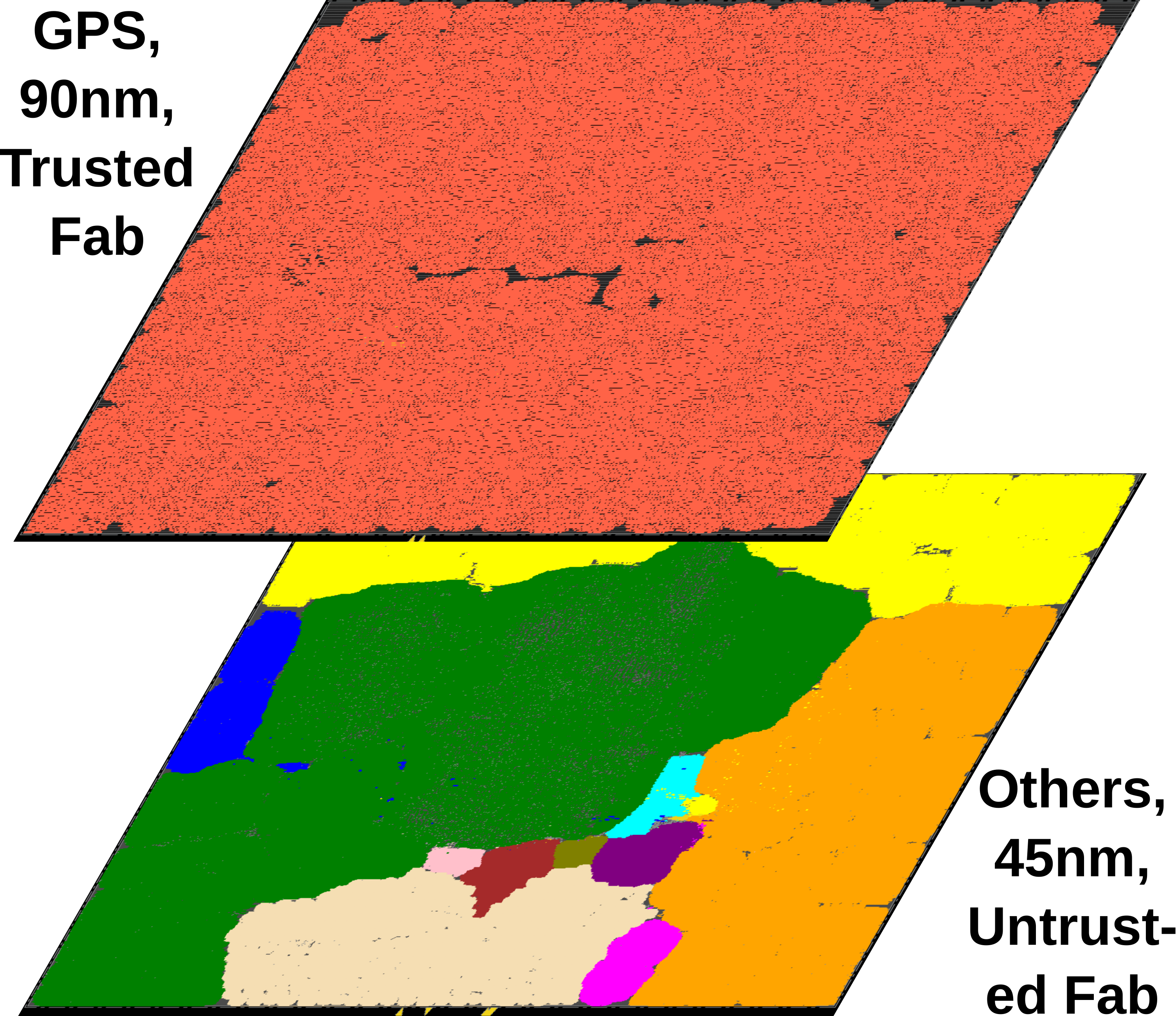}}
\smallerspacecaption
\caption{(a) Architecture of the DARPA Common Evaluation Platform~\cite{CEP_github}.
The GPS block is considered the sensitive IP; it is thus to be delegated to the trusted fabrication.
(b) Heterogeneous 3D implementation, with colors corresponding to modules in (a).}
\label{fig:CEP}
\end{figure}

Next, we study the scope for such heterogeneous 3D SM.\footnote{Another independent study is provided in
	Sec.~\ref{sec:diff_nodes_supp}.} That is, we consider a case study where the sensitive logic is moved
to a trusted fab, offering the older 90nm technology, whereas the remaining logic is delegated to an untrusted 45nm fab.  The case
study is based on multi-million-gate System-on-Chip (SoC), provided by \emph{DARPA} as a Common Evaluation
Platform (CEP)~\cite{CEP_github}.
An overview of the SoC architecture is given in Fig.~\ref{fig:CEP}(a); it is a
one-master ten-slaves system. The master is a re-engineered version of the \emph{OpenRISC} processor, called \emph{OR1200}.
This OR1200 master executes code from a 128KB static RAM (SRAM),
which we had to omit for our layout-level study, due to unavailability of memory macros in the considered 45nm library.\footnote{The memory is
	considered to be available off-chip. In any case, memory macros are not considered sensitive parts, and can thus be ignored for this
		security-focused study.}
The slaves comprise cryptographic (crypto) modules, digital signal processing (DSP) modules, and a global positioning system (GPS)
	processing module. 
The Wishbone/DBUS connect the processor master with all other blocks,
while UART provides a serial interface for
off-chip communication.

Given that all considered crypto modules are public knowledge,
	the designer (attacker) would not be interested in protecting (retrieving) the related IP.
However, the DSP modules and other parts may contain customized logic and sensitive IP worth to protect.
As discussed before, the choice of which modules/logic to protect lies solely with the design house,
as they can best judge which components require IP protection.
For our case study, we assume the GPS processing module to be that sensitive asset.

{\bf Experimental Setup:} We use the \emph{NanGate} 45nm library~\cite{nangate11} and \emph{Synopsys} 90nm
library~\cite{synopsys32nm}.
\emph{Innovus 17.1} is used for layout generation and PPA evaluation;
the setup details are further elaborated in Sec.~\ref{sec:CEP_JPEG}.
First, we synthesize the 2D baseline designs, considering the slow corners for the 45nm and 90nm nodes, respectively; the results are
given in Table~\ref{tab:CEP_orig_PPA_two_tech}.
Second, we partition the 45nm baseline design such that all logic of the GPS module can be delegated to a trusted 90nm chip.
Here we follow the design flow shown in Fig.~\ref{fig:IP_flow} in general, but there are some differences as follows:
(a)~we have to re-synthesize the GPS module for 90nm, with the same timing constraint as for the remaining modules, which is also to
simplify the clock-tree synthesis (CTS) for the individual tiers~\cite{garg13};
(b)~for layout evaluation, we revise the LIB and LEF files for the GPS tier;
(c)~we
do not undertake any additional security-centric steps (highlighted in bold in Fig.~\ref{fig:IP_flow}) as the GPS IP is fully secured
	against adversaries residing in the 45nm fab.
Finally, note that we assume supply voltages of 0.95V for
45nm and 0.9V for 90nm when generating the power numbers.\footnote{In case different parts require considerably more different voltages, level shifters should be used.
Along with technology scaling, however, the nominal voltage has
stagnated (since 90nm), which allows most modern processes to run at compatible voltages without shifters~\cite{gu2018cost}.}

\textbf{Results:} The PPA results for this different-foundries 3D implementation are given in Table~\ref{tab:CEP_45_90_PPA}.
We observe overheads of 63.09\%, 30.45\%, and 15.03\% for area, power, and delay, respectively, when compared to the 2D 45nm baseline.
However, when comparing the 3D implementation with the trusted 90nm baseline, our approach offers savings of 75.82\%, 62.31\%, and
13.1\% for area, power, and delay, respectively.
Also note that the 90nm GPS tier comprises only 15.63\% of the total number of instances, although it incurs almost the same area footprint.
In short, heterogeneous 3D integration may indeed provide benefits over a purely trusted fabrication (90nm), but it naturally cannot compete with
advanced fabrication (45nm).

\begin{table}[tb]
\centering
\scriptsize
\setlength{\tabcolsep}{0.25em}
\caption{Results for separate 2D baselines, using the \emph{NanGate} 45nm library~\cite{nangate11} and the \emph{Synopsys} 90nm
	library~\cite{synopsys32nm}.
		Layouts are DRC clean, and have a utilization of 70\%.
Area is in $\mu m^2$, power in $mW$, and delay in $ns$. 
}
\smallerspacecaption
\begin{tabular}{|c|c|c|c|c||c|c|c|c|}
\hline
\multirow{2}{*}{\textbf{Benchmark}} & \multicolumn{4}{|c||}{\textbf{45nm}} & \multicolumn{4}{|c|}{\textbf{90nm}} \\
\cline{2-9}
&
\textbf{Inst.} &
\textbf{Area} &
\textbf{Power} &
\textbf{Delay} &
\textbf{Inst.} &
\textbf{Area} &
\textbf{Power} &
\textbf{Delay} \\
\hline 
\hline
\textbf{CEP} 
&  842,685 & 1,568,893  & 429.91  & 4.41  
&  801,814 & 10,582,015 & 1,488 & 5.84 \\ \hline
\end{tabular}
\label{tab:CEP_orig_PPA_two_tech}
\end{table}

\begin{table}[tb]
\centering
\scriptsize
\setlength{\tabcolsep}{0.35em}
\caption{Results for the heterogeneous 3D implementation of Fig.~\ref{fig:CEP}(b), with the GPS module being assigned exclusively to the
	90nm tier.
	Layouts are DRC clean, and have a utilization of 70\%.
Area is in $\mu m^2$, power in $mW$, and delay in $ns$.
}
\smallerspacecaption
\begin{tabular}{|c|c|c|c|c|c|c|}
\hline
\multirow{2}{*}{\textbf{Benchmark}} 
& \multicolumn{4}{|c|}{\textbf{Area}}
& \textbf{Power}
& 
{\textbf{Delay}} \\
\cline{2-5}
& \textbf{45nm Inst.} 
& \textbf{45nm Area} 
& \textbf{90nm Inst.} 
& \textbf{90nm Area} 
&
& \\
\hline 
\hline
\textbf{CEP} & 708,858 & 1,409,486.12
& 131,342  
& 1,149,300.63
& 560.81
& 5.08 \\ \hline
\end{tabular}
\label{tab:CEP_45_90_PPA}
\smallerspacecaption
\end{table}

\subsubsection{Summary}
This approach of utilizing a trusted and another untrusted foundry for 3D integration may be inherently secure against fab-based adversaries, and
may also offer some PPA benefits over a 2D implementation using only the trusted and old node, but it also limited in practice.  That is
because one can delegate only a small fraction of the overall design to the trusted foundry (i.e., at least without incurring area cost),
which limits the scale for IP protection, and the performance and power of this 3D approach cannot compete with the
advanced but untrusted 2D node.

\subsection{Untrusted Foundries}
\label{sec:untrusted_foundry}

Engaging with (i) several untrusted foundries offering the same technology node or (ii) one untrusted foundry also holds some key
implications as follows.

First, power and performance of such ``conventional'' 3D integration can be expected to excel those of the heterogeneous scenario
above.  We note that splitting of 2D IP modules within 3D ICs has been successfully demonstrated, e.g., in \cite{
		jung17
}, albeit without hardware security in mind.  Hence, savings from the folding of IP modules may 
provide some margin for security schemes. However, we also show in detail in this work that this margin naturally depends
on the design and the measures applied for the scheme.

Second, although
IP modules can be split across tiers, which may mislead an RE attacker (a malicious end-user), both
tiers are still manufactured by some untrusted fab(s).
This fact implies that LC schemes targeting on the device level
\emph{cannot} help to protect from adversaries in those foundries.
Interestingly, there is another LC flavor emerging,
	that is the obfuscation of
interconnects~\cite{
	chen15,
	patnaik17_Camo_BEOL_ICCAD}.
We argue that obfuscation of interconnects is a natural match for F2F 3D integration---in between the two tiers, 
redistribution layers (RDLs) can be purposefully manufactured for obfuscation of the vertical interconnects (Fig.~\ref{fig:scheme}).
Doing so only requires a trustworthy BEOL facility, which is a
practical assumption given that BEOL fabrication is much
less demanding than FEOL fabrication (owing to larger pitches and less complex processing steps).
That is especially true for higher
metal layers; note that RDLs reside between the F2F bonds
which themselves are at higher layers.

Chen \emph{et al.}~\cite{chen15} consider real and dummy vias using magnesium (Mg) and magnesium-oxide (MgO), respectively, for obfuscation
of interconnects.
They demonstrate that real Mg vias oxidize quickly into MgO and, hence, can become indistinguishable from the other MgO dummy vias during
RE.
Without loss of generality, we assume our LC scheme to be based on the use of Mg/MgO vias for obfuscating the vertical interconnects
of the 3D IC.
Emerging interconnects such as those based on carbon nanotubes~\cite{
		uhlig18} may become relevant
in the future as well.

\section{Methodology for IP Protection}
\label{sec:IP_methodology}

\begin{figure}[tb]
\centering
\includegraphics[width=.9\columnwidth]{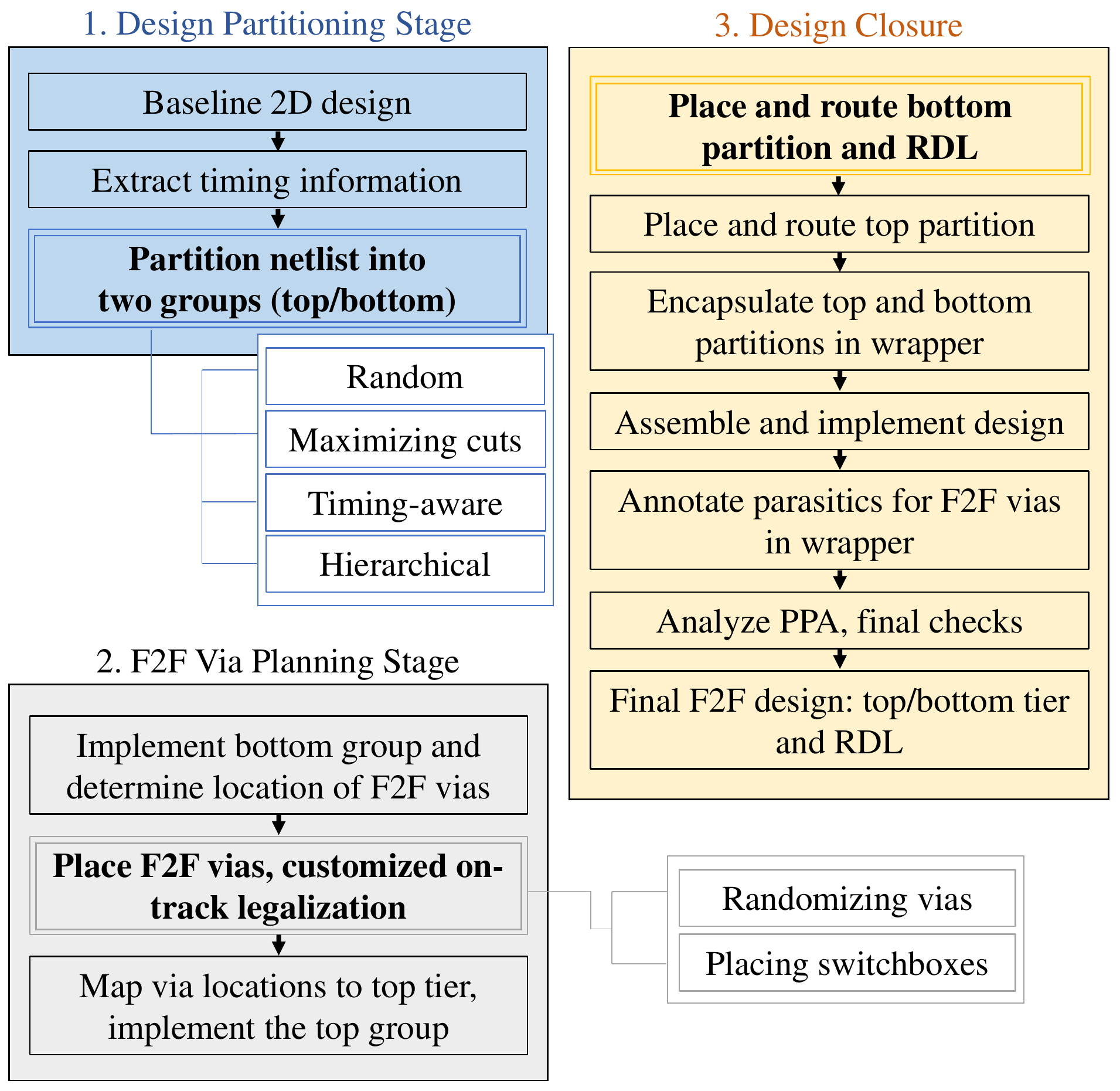}
\smallerspacecaption
\caption{Our CAD flow for F2F 3D ICs, implemented in \emph{Cadence Innovus}. Security-driven steps for IP protection are emphasized in bold.
\label{fig:IP_flow}
}
\end{figure}

Here we elaborate on our CAD and manufacturing flow for F2F 3D integration.
The CAD flow is in parts inspired by Chang \emph{et al.}~\cite{chang16}, but note that we devise and implement our customized flow, with a particular focus on IP protection
(Fig.~\ref{fig:IP_flow}).
Our flow allows a concerned designer to explore the trade-offs between PPA and \emph{cuts}, i.e., the number of F2F vertical inter-tier
connections.
Cuts are a crucial metric for the security analysis, as discussed in more detail in Sec~\ref{sec:IP_security}.

It is also important to note that we follow the call for \emph{layout anonymization}~\cite{imeson13}---we purposefully do not engage
cross-tier optimization steps, to mitigate layout-level hints on the obfuscated BEOL/RDLs.

As for the F2F process, we propose the following security-centric modification. 
The wafers for the two tiers are fabricated by one (two)
untrusted foundry (foundries) and then shipped to a trusted BEOL and stacking facility. 
This trusted facility grows the obfuscated
RDLs on top of one wafer, and continues with the regular F2F flow (i.e., flipping and bonding the second wafer on top).

\subsection{Design Partitioning}
\label{sec:design_partitioning}

After obtaining the post-routed 2D design, we partition the netlist into \emph{top} and \emph{bottom} groups, representing the 
tiers of the F2F IC.
I/O ports are created for all vertical interconnects between the two groups, representing the F2F vias.
Besides these F2F ports, we place primary I/Os at the chip boundary, as in conventional 2D designs. This is also practical for F2F
integration where TSVs are to be manufactured 
at the chip boundary for primary I/Os and 
the P/G grid.

\textbf{Random partitioning:}
A naive way for security-driven partitioning is to assign gates to the top/bottom groups randomly. 
While doing so, the number of cuts will be 
dictated by the number, type, and local inter-connectivity of gates being assigned to one group.
Since random partitioning lacks any heuristic, it 
may either result in savings or overheads for power and/or performance, depending on the design, number of vertical interconnects induced,
    and randomness itself.

\textbf{Maximizing the cut-size:}
As already indicated (and further explored in Sec~\ref{sec:IP_security}), the larger the cut size, the more difficult becomes IP piracy. 
Hence, here we seek to increase the cut size as much as reasonably possible.
First, timing reports for the 2D baseline are obtained following which gates are randomly alternated along their timing paths toward the top/bottom
groups. 
In the security-wise best case---which is also the worst case regarding power and performance---every other gate is assigned to
the top and bottom group, respectively. There, for a path with $n$ gates, $2n$ cuts are arising.
In short, the trade-off is as follows: the larger the cut size, the more resilient the design, but the higher the layout cost.
We study the impact of maximizing the cut-size in Section~\ref{sec:layout_PPA}

\textbf{Timing-aware partitioning:}
Based on the insights regarding the cost-security trade-off for random partitioning and maximizing cuts, here we seek to reduce
	layout cost while maintaining strong protection.
First, the available timing slack is determined for each gate.
Then, based on a user-defined threshold, the critical gates remain in the bottom tier, whereas all other gates are moved to the top tier.
This procedure is repeated with revised timing thresholds until an even utilization for both tiers is achieved.
Note that it is difficult for an attacker to understand whether a path in the
bottom/top group is critical or not (or even complete, for that matter).
In other words, the attacker has to tackle both groups at once and, more importantly, resolve the randomized F2F vias and the obfuscated interconnects.

We advocate this partitioning strategy, especially for any flat design. In the remainder of this work, timing-aware partitioning is
	our default strategy unless otherwise noted.

\textbf{Hierarchical partitioning:}
This strategy is applied for designs with hierarchies in the top-level module.
Inspired by~\cite{chang16},
we separate modules with a large degree of connectivity
across tiers, resulting in large numbers of cuts.
Other modules are partitioned/placed to balance the utilization of both tiers.
In short, this strategy serves to protect the IP as well as to limit layout cost for hierarchical designs.

\subsection{Planning of F2F Interconnects}
\label{sec:planning_F2F}

After placing the bottom tier, the initial locations for F2F ports are determined in the vicinity of the drivers/sinks.  Then, a
security-driven, randomized placement of F2F ports is conducted, along with customized on-track legalization.
Next, \emph{obfuscated switchboxes} are placed, and the
F2F ports are mapped to the top tier.

\textbf{Randomization:} It is easy to see that regular planning of F2F interconnects cannot be secure,
as this aligns the ports for the bottom and top tier directly.
That is, the untrusted foundry has direct access to both tiers and could simply stack them up
to recover the complete design.
Hence, we randomize the arrangement of F2F ports as follows.
(Fig.~\ref{fig:random_F2F}).  
We place additional F2F ports randomly, yet with help of on-track legalization (see below), in the top RDL.
These randomized ports are then routed through the RDLs toward the original F2F ports
connecting with the bottom tier.
In short, randomization of F2F vias is required 
to protect the design against fab-based adversaries during manufacturing.

\begin{figure}[tb]
\centering
\includegraphics[width=.52\columnwidth]{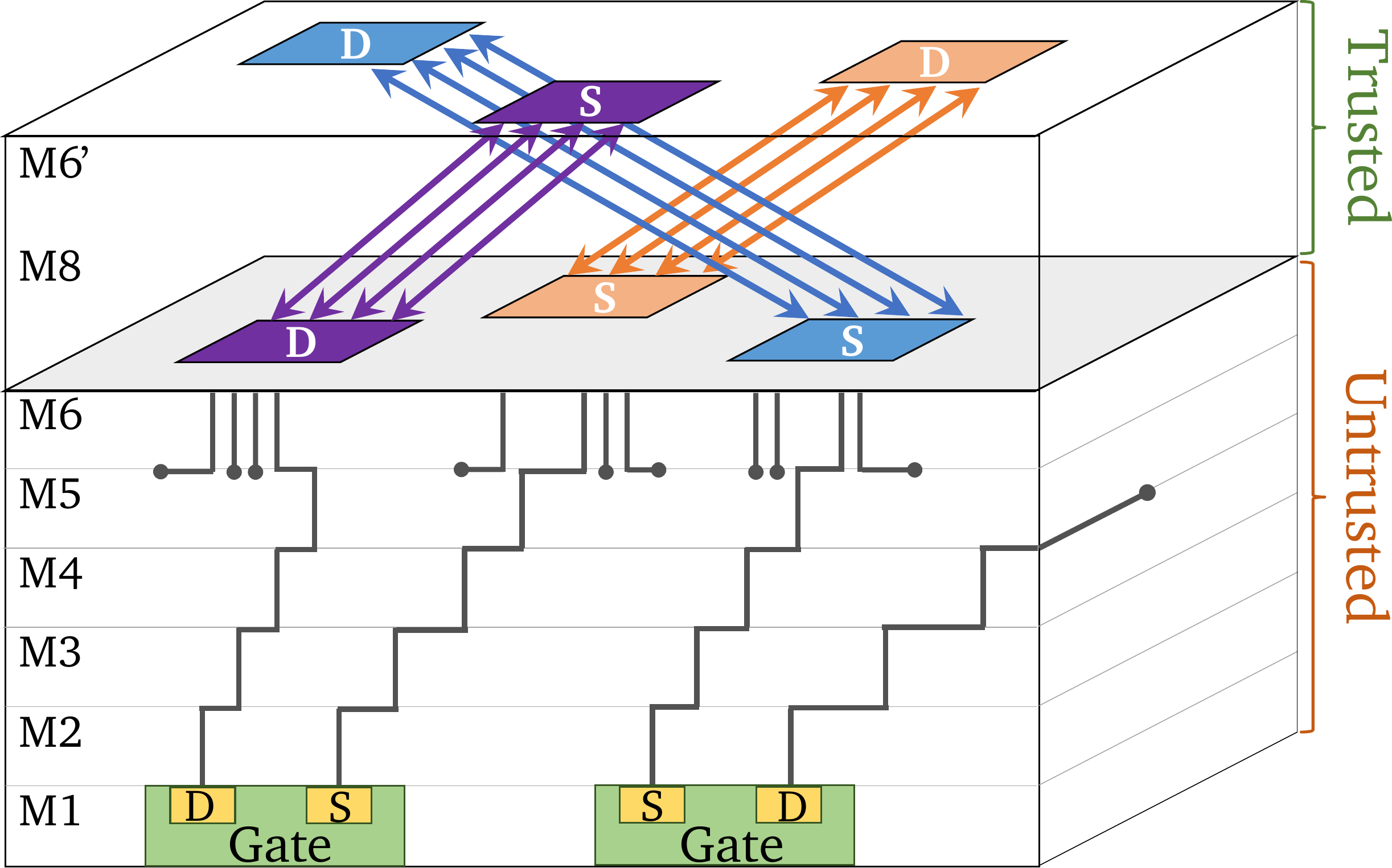}\hfill
\includegraphics[width=.45\columnwidth]{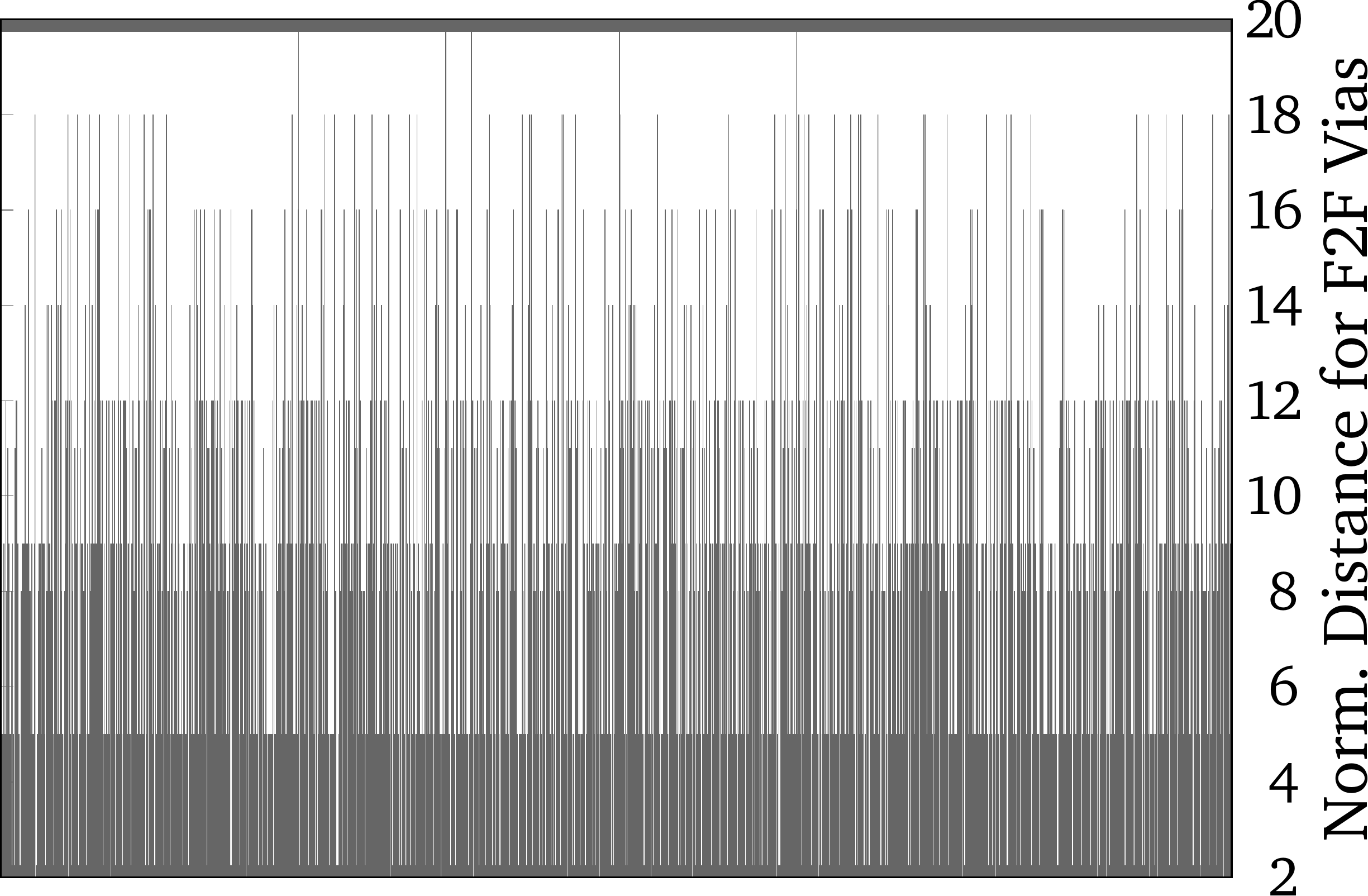}
\caption{(Left) RDL randomization for switchboxes and F2F vias.
	(Right) Normalized distances between to-be-connected F2F vias after randomization, for benchmark \emph{b17\_1}.
\label{fig:random_F2F}
}
\end{figure}

\begin{figure}[tb]
\centering
\includegraphics[width=.92\columnwidth]{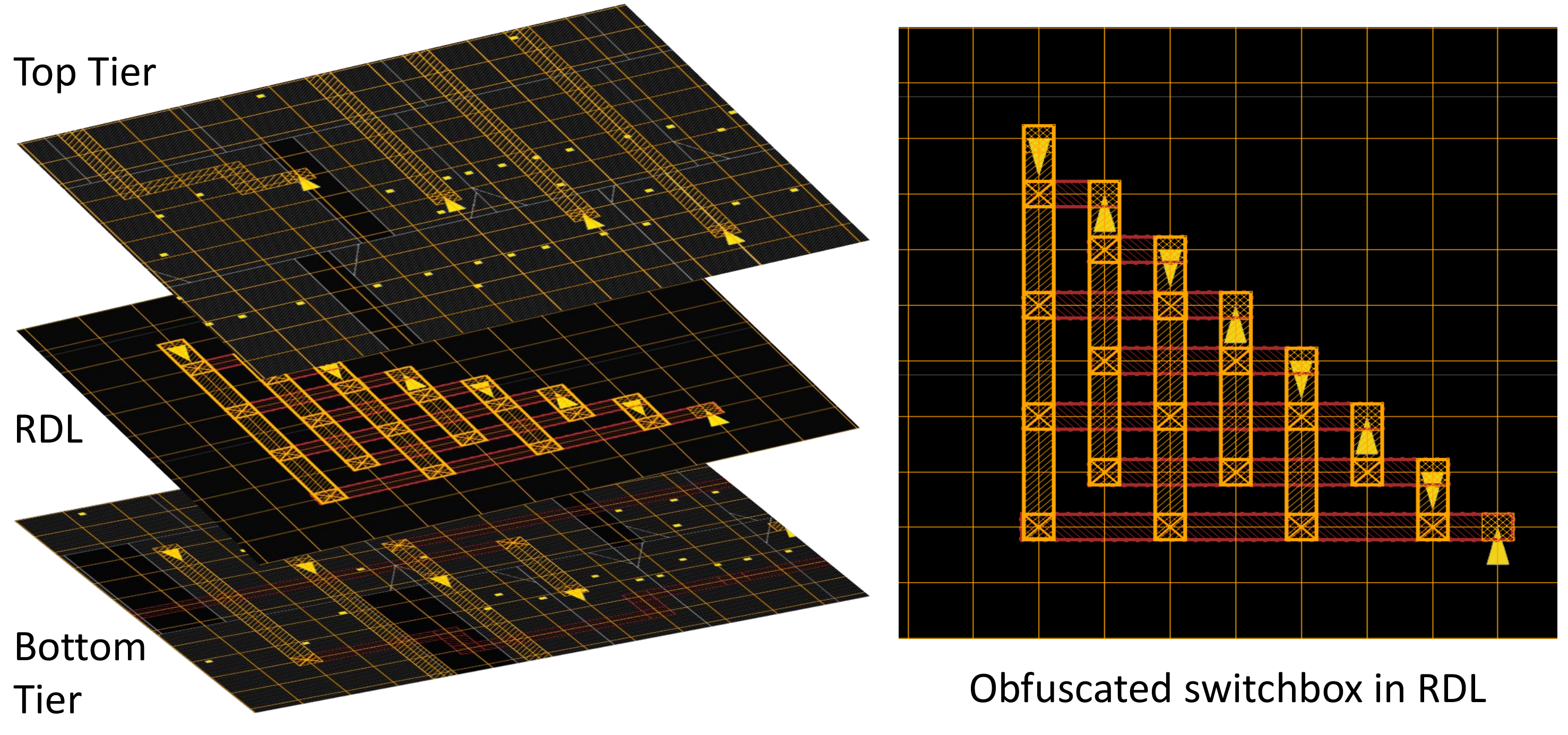}
\caption{Obfuscated switchbox, embedded in two RDL layers, exemplarily for bottom-to-top drivers.
Each driver pin (downwards triangle) can connect to any sink pin (upwards triangle).
All F2F ports are aligned with the pins of the switchbox here, for simplicity,
whereas the top-tier ports are randomized in reality.
\label{fig:switchbox}
}
\end{figure}

\textbf{Obfuscated switchboxes:}
To further protect against RE attacks  from malicious end-users, we obfuscate the
connectivity in the RDLs,
using
a customized switchbox (Fig.~\ref{fig:switchbox}).
This switchbox allows stealthy one-to-one mapping of four drivers to four sinks.
The essence of the switchbox is the use of Mg/MgO vias
(recall Sec.~\ref{sec:untrusted_foundry}), to cloak which driver
connects to which sink.
The pins of the switchbox represent the F2F ports.
The pins are aligned with
the routing tracks to enable proper
utilization of routing resources.
For randomization, the additional ports
connecting with the top tier are used for rerouting during design closure.

\textbf{On-track legalization:}
Each F2F port is moved inside the core boundary, toward the center point defined by all instances connected with this port.
Next, we obtain the closest and still-unoccupied on-track location for actual placement.
If need be, we step-wise increase the search radius considering a user-defined threshold.

\subsection{Design Closure}
\label{sec:closure}

After the F2F via planning stage, both tiers are placed and routed separately, independent of each other.
For sequential designs, we conduct CTS on both tiers independently, as suggested in~\cite{garg13}.

Recall that we do not engage in any cross-tier optimization, on purpose, to anonymize the individual tiers
from each other. However, we apply intra-tier optimization.
While routing the bottom tier, we also route the randomized
and obfuscated RDL with their switchboxes. 
Next, we encapsulate 
the top and bottom partitions in a \emph{wrapper netlist}, and we assemble and implement the design followed by generating a 
Standard Parasitic Exchange Format (SPEF)
file that captures the RC parasitics
of the F2F vias (modeled as regular vias, see below).
Finally, we perform DRC checks, evaluate the PPA, and stream out separate DEF files for the top/bottom tiers and the RDL.

\section{Results for IP Protection}
\label{sec:IP_results}

\subsection{Experimental Setup}
\label{sec:IP_setup}

\textbf{Implementation and layout evaluation:}
Since there are no commercial tools available yet for (F2F) 3D ICs, 
we implement our CAD flow within \emph{Cadence Innovus 17.1}, using  custom \emph{TCL} and \emph{Python} scripts.
Our implementation imposes negligible design runtime overheads.
We use the \emph{NanGate} 45nm library~\cite{nangate11} for our experiments, with six metal layers for the
baseline 2D setup and six layers each for the top and bottom tier in the F2F setup.
The RDL comprises four duplicated layers of M8,
from which two are used for embedding the obfuscated switchboxes, and two are used for randomizing the routing.
F2F vias are modeled as M6 vias;
while this is an optimistic assumption, for now,
F2F technology scaling can be expected to reach such dimensions.
The PPA analysis is conducted for the slow process corner, using \emph{CCS} libraries at 0.95V. 
For power analysis, we assume a switching activity of 0.2 for
all primary inputs.
We ensure that the layouts are free of any congestion, by choosing appropriate utilization rates
for the 2D baselines.
This is essential to prevent any 
possible congestion to be carried forward in our 3D flow.
All experiments are carried out on an
Intel Xeon E5-4660 @ 2.2 GHz with \emph{CentOS 6.9}. For \emph{Cadence Innovus}, up to 16 cores are allocated.

\textbf{Setup for security evaluation:}
Since we promote 3D SM, regular proximity attacks such as~\cite{rajendran13_split,wang18_SM
} cannot be applied.
Thus, we propose and publicly release
a novel attack against 3D SM~\cite{webinterface},
also accounting for the RDL obfuscation underlying in our scheme;
see also Sec.~\ref{sec:IP_security}.
Attacks on our protected layouts are evaluated by commonly used metrics, i.e., the \emph{correct connection rate (CCR)}, \emph{percentage of
netlist recovery (PNR)}~\cite{patnaik18_SM_ASPDAC}, and \emph{Hamming
distance (HD)}.  HD is calculated using \emph{Synopsys VCS} with 1,000,000 test patterns.
As for
SAT-based RE attacks, we leverage the tool provided by~\cite{subramanyan15}, with
the related time-out set to 72 hours.

\textbf{Designs:}
The commonly considered benchmarks from the \emph{ISCAS-85} and \emph{ITC-99} suites are used for layout and security analysis.
In addition, we also use two SoC benchmarks: the DARPA CEP~\cite{CEP_github} and the JPEG OpenCores design~\cite{Opencores}.

\subsection{Security-Driven Layout Evaluation}
\label{sec:layout_PPA}

\begin{figure}[t]
\centering
\includegraphics[width=.99\columnwidth]{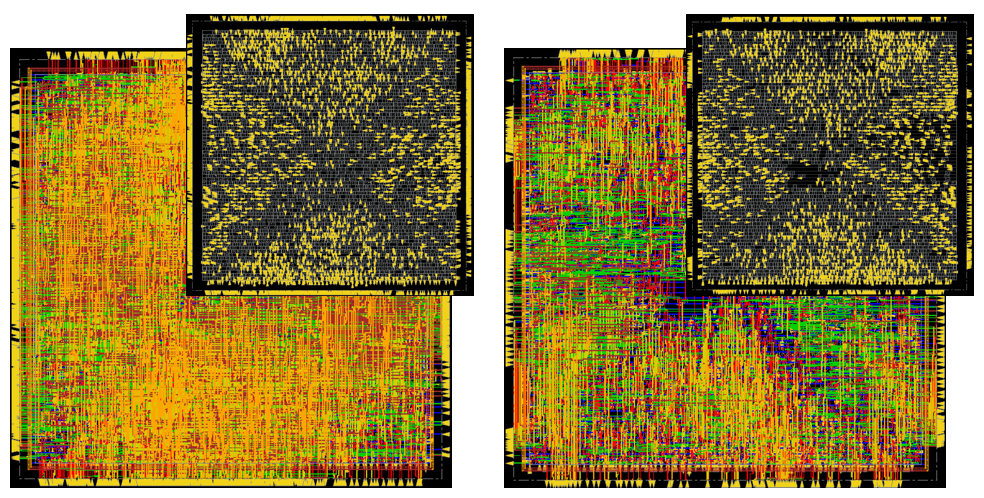}
\smallerspacecaption
\caption{Layout snapshots of bottom/top tier (left/right) for \emph{b22}. The insets show the corresponding F2F vias.
\label{fig:IP_die_images}
}
\end{figure}

Our flow allows to trade-off PPA and cuts; the latter dictates the resilience against IP piracy both during and
after manufacturing.
Figure~\ref{fig:IP_die_images} showcases the layout images for
benchmark \emph{b22}.

\begin{figure*}[ht]
\centering
\captionsetup[subfigure]{labelformat=empty}
\subfloat[]{\includegraphics[width=.99\textwidth]{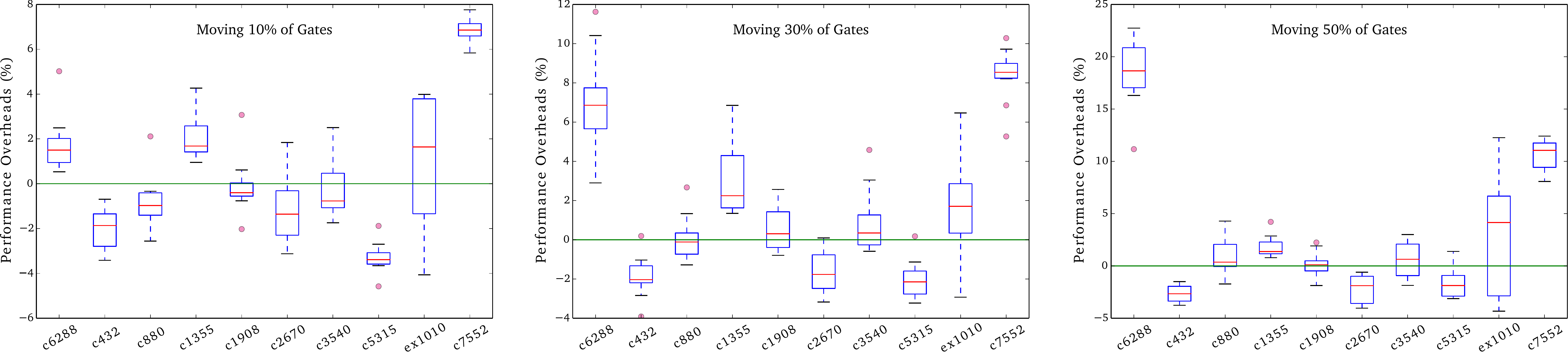}}\\
\smallerspacecaption
\smallerspacecaption
\smallerspacecaption
\smallerspacecaption
\subfloat[]{\includegraphics[width=.99\textwidth]{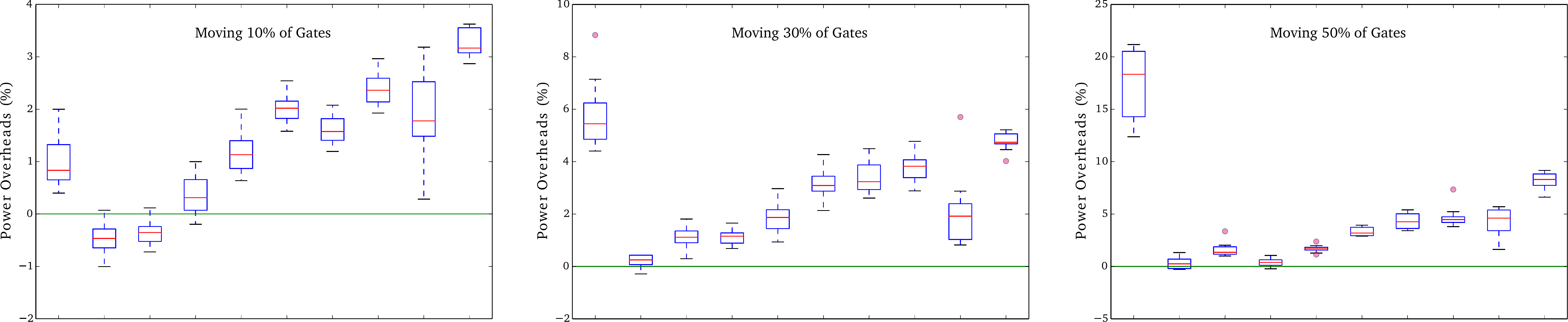}}
\smallerspacecaption
\smallerspacecaption
\caption{Impact of randomly assigning gates on performance (top) and power (bottom). Each boxplot represents ten runs.
Note that the same benchmarks are applied for 
the top and bottom plots; benchmark labels are accordingly placed between those plots.
\label{fig:PPA_MC}
}
\end{figure*}

\begin{figure}[htb]
\centering
\includegraphics[width=.99\columnwidth]
{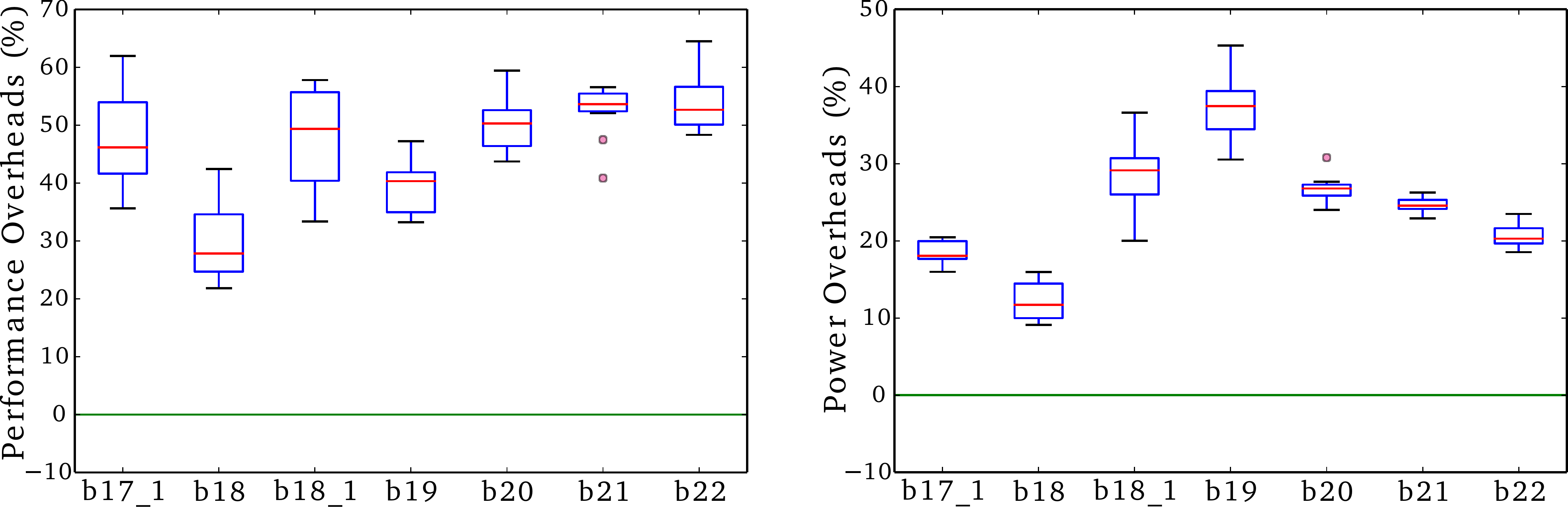}
\smallerspacecaption
\caption{Layout cost for maximizing cuts, with 35--50\% of the gates moved, and with obfuscated switchboxes and F2F randomization being applied. Each boxplot represents ten runs.}
\smallerspacecaption
\label{fig:MC_plus_pin_random_cross}
\end{figure}

\begin{figure}[htb]
\centering
\includegraphics[width=.99\columnwidth]
{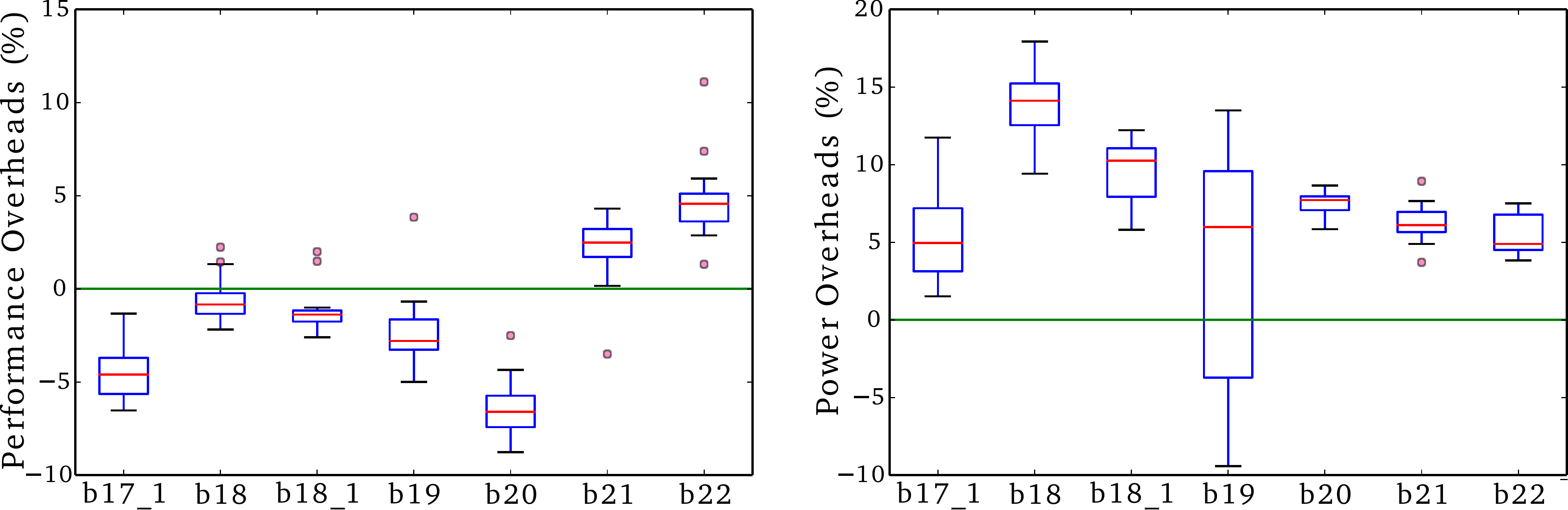}
\smallerspacecaption
\caption{Performance, power cost for timing-aware setup with obfuscated switchboxes and F2F randomization. Each box represents ten runs.}
\smallerspacecaption
\label{fig:TA_plus_pin_random_cross}
\end{figure}

\subsubsection{Random Partitioning and Maximizing the Cut-Size}
Initially we study random partitioning of gates, by moving them randomly
from the bottom to the top group
in steps of 10\%, up to 50\%.
As the strategy is randomized, we perform ten runs for each benchmark for any given percentage of gates to move.
The resulting power and performance distributions are illustrated
in Fig.~\ref{fig:PPA_MC}.

Interestingly, even for the security-wise best case of randomly moving 50\% of the gates,
some runs still provide better power and/or performance than the 2D baseline.
The savings in performance can be attributed to the fact that, when splitting the design across the vertical dimension, we can obtain a
	reduction in wirelength, which helps to improve timing.
We note that these improvements, on average, come at some expense of power, with related overheads in the range of 0--7\% for lifting/moving 50\% of gates.

While this demonstrates the potential for naive random partitioning, it is important
to note that this finding only holds true as long as we refrain from randomizing the F2F ports
and from using the obfuscated switchboxes for these experiments.
In fact, once we seek to maximize the cuts, along with randomization of F2F ports and use of switchboxes, larger \emph{ITC-99} benchmarks such as \emph{b18\_1} incur considerable overheads of
up to 60\% (Fig.~\ref{fig:MC_plus_pin_random_cross}).
Here we also observe that large cut-sizes lead to an increase in routing 
congestion and total wirelength, thereby further increasing the total capacitance of the design.
This offsets the performance benefits which regular, security-oblivious 3D integration can  be expected to achieve~\cite{ku18}.

In short, although these strategies offer strong resilience, a more aggressive
PPA-security trade-off may be desired.

\subsubsection{Timing-Aware Partitioning}
This setup tackles that need for achieving 
security while maintaining reasonable PPA cost. 
We observe that even for larger \emph{ITC-99} benchmarks such as \emph{b18\_1} and \emph{b19} (Fig.~\ref{fig:TA_plus_pin_random_cross}),
there are some
benefits when comparing the secure 3D designs to 
their 2D baseline.
As explained in Sec.~\ref{sec:design_partitioning}, since the most timing-critical gates are constrained to one tier, we induce
significantly less cuts along the timing paths for the 3D design. 
For example, we observe a reduction of about 60\% in timing-path cuts for \emph{ITC-99} benchmarks \emph{b18\_1} and \emph{b19} when compared to random partitioning.
To demonstrate the
security implication of this setup, we plot the normalized distances between to-be-connected F2F vias in
Fig.~\ref{fig:random_F2F}. This figure shows a wide variation across the inter-tier nets, whereas for regular, unprotected F2F stacking the
distances would be all zero.
Overall, the choice of partitioning lies with the designer, which she/he can trade-off considering security and PPA cost, but
timing-aware partitioning should be considered first, i.e., at least for non-hierarchical designs.

\subsubsection{Case Study on CEP
and JPEG}
\label{sec:CEP_JPEG}
Besides the well-known benchmarks considered above, we also
conduct case studies on two ``real-world netlists,'' the DARPA CEP~\cite{CEP_github} and
the JPEG OpenCores design~\cite{Opencores}.
Thus, we also demonstrate our secure end-to-end CAD flow for practical 3D ICs.

\textbf{Setup:}
The utilization is set to 70\%
and 60\% for CEP and JPEG, respectively, which
ensures that the 2D baseline designs are devoid of any congestion.
We use the \emph{NanGate} 45nm library~\cite{nangate11}.
F2F vias are modeled as M10 vias.
All 2D and 3D designs operate at iso-performance, with a timing
constraint of 5$ns$ (i.e., at 200 MHz).
Further details are the same as in Sec.~\ref{sec:IP_setup}.

\textbf{Results:} The results for the 2D baseline and secure 3D designs are provided in Table~\ref{tab:PPA_CEP_JPEG_both_45}.
Regarding the footprint area/die outlines, both the secure 3D designs provide savings over their 2D baselines, namely 49.3\% for CEP and
42\% for JPEG.
Regarding instance counts, we observe some overheads for both the secure 3D designs;
as we do not apply any cross-tier and/or post-partitioning optimization, there is less leverage to reduce instance counts for the tools.
Again, for regular, security-oblivious 3D F2F integration,
one would expect savings/reductions in 
both wirelength and instance count, which ultimately also enables power savings~\cite{ku18}.

For our security-driven 3D flow, it depends on various aspects
whether there are power/performance savings or overheads.
First, recall that we randomize F2F vias and leverage obfuscated switchboxes
(to deter fab-based adversaries and malicious end-users, respectively), which tends to increases wirelength, and thereby the driver
strengths and/or buffer counts.
Second, the designs have some impact by themselves.
For example, for JPEG we note 8.72\% higher power consumption, whereas for CEP we note a 6.49\% power reduction.
Third, partitioning plays an important role as well, as already discussed.
Since CEP and JPEG are both hierarchical designs, we apply hierarchical partitioning here, which fully protects the system-level IP
orchestration and any glue logic. To further protect individual modules, we can split them up across the two tiers; which
modules to select and how to split is the designer's decision, also depending on the nature of the modules and the overall design~\cite{jung17}.
Toward this end, we also performed an experiment
on CEP
where individual modules were partitioned, resulting in 20,863 F2F vias (3.25$\times$ than those reported in
Table~\ref{tab:PPA_CEP_JPEG_both_45}). 
We maintain that both tiers are DRC-clean and
free of congestion. We observe power and timing overheads of 
9.59\% and 13.64\%, respectively,
	which implies that this 3D design can operate at around 176 MHz.
	Running the overall chip at this frequency also ensures that there is no loss in system functionality.
Finally, note that the number of cuts we obtain here indicates strong resilience of the 3D designs; we discuss this further in
Sec.~\ref{sec:IP_security}.

\begin{table}[tb]
\centering
\scriptsize
\setlength{\tabcolsep}{0.5em}
\caption{Comparison between 2D baseline designs and 
their secure 3D F2F counterparts 
of CEP~\cite{CEP_github} and JPEG~\cite{Opencores}.
All results are for iso-performance (5$ns$).
Wirelengths in 3D are subject to
randomization of F2F vias.}
\smallerspacecaption
\begin{tabular}{|c|c|c|c|c|c|c|c|c|}
\hline
\multirow{2}{*}{\textbf{Metrics}} 
& \multicolumn{2}{|c|}{\textbf{CEP}} 
& \multicolumn{2}{|c|}{\textbf{JPEG}} \\
\cline{2-5}
&
\textbf{2D} &
\textbf{3D} &
\textbf{2D} &
\textbf{3D} \\
\hline 
\hline
Footprint ($\mu m^2$) & 2,332,429 & 1,180,807  & 1,317,041 & 763,945 \\ \hline
F2F Via Count & 0 & 6,447 & 0 
& 6,707 \\ \hline
Total Cell Count & 852,496 & 859,207 & 497,666 
& 520,374 \\ \hline
Total Wirelength ($m$) & 26.62 & 23.08 & 8.69 
& 9.58 \\ \hline
Total Power ($mW$) & 417.8 & 390.7 & 204.1 
& 221.9 \\ \hline
\end{tabular}
\label{tab:PPA_CEP_JPEG_both_45}
\smallerspacecaption
\end{table}

\subsection{Comparison with Prior Art}
\label{sec:PPA_comparison_SM_LC}

\textbf{LC schemes:}
Among others, threshold-voltage-dependent LC is gaining traction. Although promising concerning resilience,
the
PPA cost are considerable.  For example, Akkaya \emph{et al.}~\cite{akkaya18} report overheads of 9.2$\times$,
6.6$\times$, and 3.3$\times$ for PPA, respectively, when compared to conventional 2-input NAND gate.
Nirmala~\emph{et al.}~\cite{nirmala16} report 11.2$\times$ and 10.5$\times$ cost for power and area, respectively.
Besides, for interconnects camouflaging, Patnaik~\emph{et al.}~\cite{patnaik17_Camo_BEOL_ICCAD} report PPA overheads of 4.9\%, 31.2\%, and
25\% for \emph{ITC-99} benchmark \emph{b17} at 60\% LC.
When compared to these schemes, we can provide
significantly better PPA (except for~\cite{patnaik17_Camo_BEOL_ICCAD} concerning power).

Regarding prior art on 3D LC~\cite{yan17_camo,gu2018cost}, recall that they require a trusted FEOL facility; hence, their schemes are
not directly comparable to ours. 
Also, at the time of writing, their libraries and protected designs were not available to us 
for a detailed study.
Moreover, for~\cite{gu2018cost}, the authors leverage regular 2D LC schemes while using
different technology nodes. Depending on the particular node and LC scheme, this may induce large PPA overheads, and
technology-heterogeneous 3D integration may hold further complications~\cite{peng17,garg13}.
Recall that these concerns were our main
motivation to advocate the use of uniform/same technologies and camouflaging of vertical interconnects.

\textbf{SM schemes:}
In Table~\ref{tab:sec-comp2}, we compare with studies on 2D SM.
Overall, the placement-centric techniques by Wang \emph{et al.}~\cite{wang18_SM}
are competitive 
concerning
power and performance. However, as always the case for regular SM,
      they
can only avert fab-based adversaries,
\emph{but not malicious end-users}.

\begin{table*}[tb]
\centering
\footnotesize
\setlength{\tabcolsep}{1mm}
\caption{PPA cost comparison with 2D SM protection schemes. 
Numbers are in \% and quoted from the 
respective publications.}
\smallerspacecaption
\begin{tabular}{|*{16}{c|}}
\hline
\multirow{2}{*}{\textbf{Benchmark}}
& \multicolumn{3}{|c|}{\textbf{BEOL+Physical~\cite{wang18_SM}}} 
& \multicolumn{3}{|c|}{\textbf{Logic+Physical~\cite{wang18_SM}}} 
& \multicolumn{3}{|c|}{\textbf{Logic+Logic~\cite{wang18_SM}}} 
& \multicolumn{3}{|c|}{\textbf{Concerted Lifting~\cite{patnaik18_SM_ASPDAC}}} 
& \multicolumn{3}{|c|}{\textbf{Proposed with Random Partitioning}} \\
\cline{2-16}
& \textbf{Area} & \textbf{Power} & \textbf{Delay}
& \textbf{Area} & \textbf{Power} & \textbf{Delay}
& \textbf{Area} & \textbf{Power} & \textbf{Delay}
& \textbf{Area} & \textbf{Power} & \textbf{Delay}
& \textbf{Area$^*$} & \textbf{Power} & \textbf{Delay}
\\ \hline
c432 & N/A & 0.17 & 0.49 &
N/A & 0.44 & 0.24 &
N/A & 0.17 & 0.21 &
7.7 & 13.1 & 11.6 &
-50 & -2.66 & 0.31
 \\ \hline
 
c880 & N/A & 0.25 & 0.05 &
N/A & 0.35 & 0.03 &
N/A & -0.05 & -0.09 &
0 & 12.1 & 19.9 &
-50 & 0.97 & 1.6
 \\ \hline
 
c1355 & N/A & 0.52 & 0.57 &
N/A & 0.75 & 0.42 &
N/A & 0.03 & 0.01 &
0 & 12.2 & 21.3 &
-50 & 1.83 & 0.38
\\ \hline

c1908 & N/A & 1.1 & 1.3 &
N/A & 1.1 & 0.23 &
N/A & 0.45 & 0.39 &
7.7 & 14.6 & 18.9 &
-50 & 0.11 & 1.69
\\ \hline

c2670 & N/A & 0.29 & 0.27 &
N/A & 0.29 & 0.27 &
N/A & 0.05 & 0.03 &
7.7 & 10 & 12 &
-50 & -2.18 & 3.32
\\ \hline

c3540 & N/A & 0.53 & 0.28 &
N/A & 0.36 & 0.02 &
N/A & 0.14 & -0.02 &
7.7 & 5 & 2.8 &
-50 & 0.59 & 4.32
\\ \hline

c5315 & N/A & 0.19 & -0.01 &
N/A & 0.67 & 0.08 &
N/A & 0.29 & -0.01 &
7.7 & 7.9 & 16.9 &
-50 & -1.66 & 4.73
\\ \hline

c6288 & N/A & 0.29 & 0.19 &
N/A & 0 & 0 &
N/A & 0.1 & 0.67 &
27.3 & 12.3 & 15.7 &
-50 & 10.43 & 10.21
\\ \hline

c7552 & N/A & 0.28 & -0.36 &
N/A & 0.35 & -0.05 &
N/A & 0.56 & 1.77 &
16.7 & 9.3 & 15.7 &
-50 & 10.57 & 8.21
\\ \hline

\textbf{Average} & N/A & 0.4 & 0.31 &
N/A & 0.48 & 0.14 &
N/A & 0.19 & 0.33 & 
9.2 & 10.7 & 15 &
-50 & 2 & 3.86
\\ \hline

\end{tabular}
\\ \footnotesize
$^*$Following the standard practice for 3D studies, we report on area by considering individual die outlines.
In~\cite{patnaik18_SM_ASPDAC}, area is reported in terms of die outlines as well.
\label{tab:sec-comp2}
\end{table*}

In Table~\ref{tab:comparison_with_2.5D}, we compare with the security-driven 2.5D integration scheme by Xie \emph{et al.}~\cite{xie17}.
Their work is relevant as they
propose a similar notion of security based on cut sizes. For the benchmarks the authors
considered, we obtain on average 53\% more cuts in our scheme. (For our cut sizes on larger benchmarks, refer to
		Table~\ref{tab:security_analysis}).
Regarding
PPA, we observe significantly lower costs than~\cite{xie17}.\footnote{\label{fn:die_outlines}
Concerning area, note that we report on die outlines, which is
standard practice for 3D studies.
Accordingly, for our result of -50\%, the 3D IC and the 2D baseline require the same total silicon area. In other words, we incur 0\%
additional area cost.
While Xie \emph{et al.} report on additional area cost,
they omit that their scheme requires
an interposer which---being at least as large as the chips stacked onto it---incurs $\geq$100\% cost.
Still, when only comprising metal layers, we acknowledge that an interposer is less expensive than regular chips.}
Besides, as with regular SM, their 2.5D scheme is 
\emph{not} inherently
   resilient against malicious end-users, but our 3D scheme is.

\begin{table}[tb]
\centering
\scriptsize
\setlength{\tabcolsep}{0.2em}
\caption{Comparison with 2.5D scheme 
of~\cite{xie17}. PPA is in contrast to a 2D baseline, numbers are in \%.
See also Footnote~\ref{fn:die_outlines} on area cost.
}
\smallerspacecaption
\begin{tabular}{|c|c|c|c|c|c|c|c|c|}
\hline
\multirow{2}{*}{\textbf{Benchmark}} 
& \multicolumn{4}{|c|}{\textbf{Xie \emph{et al.}~\cite{xie17} (SC+SP)}} 
& \multicolumn{4}{|c|}{\textbf{Proposed with Random Partitioning}} \\
\cline{2-9}
 &
 \textbf{Cut Size} &
 \textbf{Area} &
 \textbf{Power} &
 \textbf{Delay} &
 \textbf{Cut Size} &
 \textbf{Area} &
 \textbf{Power} &
 \textbf{Delay} \\
 \hline \hline
c432 &  
130 & 1  & 17.6  & 5.9  & 134  & -50 (0) & -2.66 & 0.31 \\ \hline
c880 &  
141 & 0  & 29.4  & 10  & 138  & -50 (0) & 0.97 & 1.6 \\ \hline
c1355 &  
130 & 0  & 17.6  & 17.6 & 91  & -50 (0) & 1.83 & 0.38 \\ 
\hline
c1908 &  
132 & 1  & 11.8  & 29.4 & 149  & -50 (0) & 0.11 & 1.69 \\
\hline
c2670 &  
152 & 0  & 11.8  & 5.9 & 154  & -50 (0) & -2.18 & 3.32 \\
\hline
c3540 &  
133 & 0  & 5.9  & 5.9 & 349  & -50 (0) & 0.59 & 4.32 \\
\hline
c7552 &  
157 & 1  & 1  & 5.9 & 477  & -50 (0) & 10.57 & 8.21 \\
\hline
\textbf{Average} &  
139 & 0.4  & 13.6  & 11.5 & 213  & -50 (0) & 1.32 & 2.83 \\
\hline
\end{tabular}
\label{tab:comparison_with_2.5D}
\smallerspacecaption
\end{table}

\subsection{Security Analysis and Attacks}
\label{sec:IP_security}

\subsubsection{Our Proximity Attack for 3D SM}
To the best of our knowledge, there is no attack yet in the literature which can account for 3D SM in the context of IP piracy.
Hence, we propose and implement such an attack, with a focus on one untrusted foundry (or two colluding foundries) and our RDL obfuscation. We
provide this attack as a public release in~\cite{webinterface}.

We assume that the attacker holds the layout files for the top and bottom tier,
but, residing in the untrusted fab, she/he has 
no access to the trusted RDL.\footnote{We discuss the implications for malicious end-users being
able to access the obfuscated RDL further below.}
Although she/he understands how many drivers are connecting from the bottom to the top tier and vice versa, 
she/he does not know which driver connects to which sink, given the randomization of F2F vias.
Recall that, we do not engage in cross-tier optimization, to mitigate any layout-level hints.
Let us assume there are $d_{bot}$ drivers in the bottom and, independently, $d_{top}$ drivers in the top tier.
Since we do not allow for fan-outs within the RDL (this would occupy more F2F vias than necessary), there are only one-to-one
mappings---this results in $d_{bot}! \times d_{top}!$ possible netlists.
Once switchboxes are used, however,
the attacker can tackle groups of four drivers/sinks at once.
Still, she/he has to resolve (a)~which four top-tier drivers are connected to which four bottom-tier sinks and vice versa, and (b) the
connectivity within the obfuscated switchboxes.
For those cases, there are
$4!\times \left(\left(1/4 \times d_{bot}\right)! \times \left(1/4 \times d_{top}\right)!\right)$ possible netlists remaining.
Next, we outline the corresponding heuristics at the heart of our attack.

\begin{enumerate}

\item \emph{Unique mappings:} Any driver in the bottom/top tier will feed only one sink in the top/bottom tier.
Hence, an attacker will reconnect drivers and sinks individually.
Moreover, she/he can identify all primary I/Os as they are
implemented using wirebonds or TSVs, not randomized F2F vias.

\item \emph{Layout hints:} Although the F2F vias are randomized, the attacker may try to correlate the proximity and
orientation of F2F vias with their corresponding RDL connectivity.
Toward this end, she/he can also investigate the routing toward 
the switchbox ports.
Moreover, recalling the practical threat model, the attacker may be able to identify some known IP and 
confine the related sets of candidate F2F interconnects accordingly.
Our attack is generic and can account for those scenarios, by keeping track of the candidate F2F pairings considered by the attacker.

\item \emph{Combinatorial loops:} Both tiers and thus all active components are available to the attacker, hence she/he can readily exclude
those F2F connections inducing combinatorial loops.

\end{enumerate}

\begin{figure*}[tb]
\centering
\includegraphics[width=\textwidth]{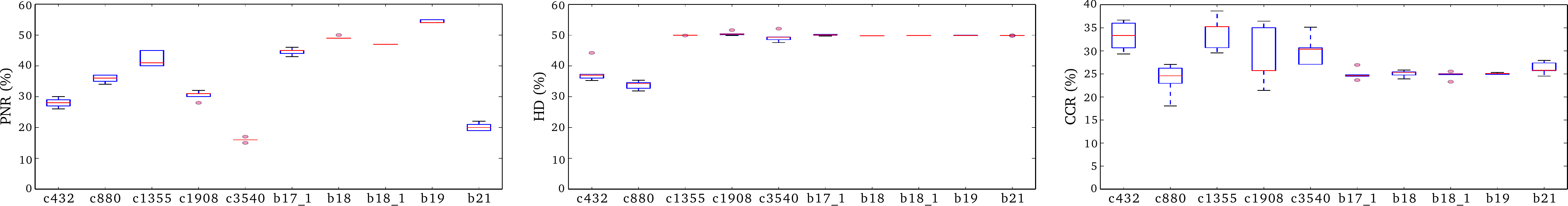}
\caption{Percentage of netlist recovery (PNR)~\cite{patnaik18_SM_ASPDAC}, Hamming distance (HD), and correct connection rates (CCR), when the benchmarks are subjected to our 3D SM proximity attack~\cite{webinterface}.
Each box represents ten	runs.
\label{fig:3D_proximity_attack}
}
\end{figure*}

The results in Table~\ref{tab:security_analysis} indicate the efficiency of our proposed proximity attack (especially over the SAT-based
attack~\cite{subramanyan15}, see below for that scenario).
Here we assume that the attacker is able 
to infer all the driver-sink
pairings for the switchboxes correctly;
only the obfuscation within switchboxes remain 
to be attacked.
This is a \emph{strong} assumption and, hence, rendering our evaluation conservative.
In fact, this scenario can be considered as an optimal proximity attack, as
for all F2F connections the correct one is always among the considered candidates.
With regards to CCR, PNR, and HD for the recovered netlists, our protection scheme can be considered as reasonably secure
(Fig.~\ref{fig:3D_proximity_attack}).
Although PNR, which represents the degree of similarity between the original and the recovered netlist~\cite{patnaik18_SM_ASPDAC}, is 
around 30\% or more for most benchmarks, HD approaches the ideal value of 50\% for most benchmarks. In other words, although our attack
can correctly recover some parts of the design, the overall functionality still remains obscured.

\begin{table}[tb]
\centering
\scriptsize
\setlength{\tabcolsep}{0.5em}
\caption{Cut sizes and average attack runtimes. Time-out `t-o' is 72 hours.
}
\smallerspacecaption
\begin{tabular}{|c|c|c|c|c|
}
\hline
\multirow{2}{*}{\textbf{Benchmark}} 
& \multicolumn{2}{|c|}{\textbf{Cut Sizes}}
& \textbf{SAT Attack~\cite{subramanyan15}}
& 
{\textbf{Proposed Attack}} \\
\cline{2-5}
 &
 \textbf{Random} &
 \textbf{Timing-Aware} &
 \textbf{Runtime (Min.)} &
 \textbf{Runtime (Sec.)} \\
 \hline \hline
c432 & 134 & 56 & 624
& 0.004 \\ \hline
c880 & 138 & 53 & 642
& 0.003 \\ \hline
c1355 & 91 & 37 & 492
& 0.07 \\ \hline
c3540 & 349 & 97 & 948
& 8.73 \\ \hline
b17\_1 & 6,650 & 2,482 & t-o
& 0.25 \\ \hline
b18 & 15,974 & 6,906 & t-o
& 113.51 \\ \hline
b18\_1 & 16,706 & 6,616 & t-o
& 4.62 \\ \hline
b19 & 33,417 & 13,142 & t-o
& 0.53 \\ \hline
\end{tabular}
\label{tab:security_analysis}
\smallerspacecaption
\end{table}

\subsubsection{SAT-Based Attacks}
After manufacturing, the attacker can readily understand which four drivers/sinks are connected through the switchboxes, but she/he still has
to resolve the obfuscation within the switchboxes themselves.
The attacker may now leverage a working copy as an oracle and launch a SAT attack.
Toward that end, we employ the attack proposed in~\cite{subramanyan15}, and we model the problem using multiplexers.
Empirical results are given in Table~\ref{tab:security_analysis}. 
As expected, the SAT-based attack succeeds for smaller designs but runs
into time-out for larger designs. 
This finding is also consistent with those reported by Xie \emph{et al.}~\cite{xie17} for their security-driven
2.5D scheme, which has a security notion similar to our work.

\section{Methodology for Trojan Prevention}
\label{sec:HT_methodology}

So far we have leveraged the potential of 3D integration for IP protection. 
As elucidated earlier, insertion of hardware Trojans (HTs) by an untrustworthy foundry is another concerning threat.
Security schemes like SM can hinder those adversaries from obtaining and fully understanding the netlist;
hence, the adversaries may fail to insert HTs at particular targeted locations.
However, there are many other parties in an IC supply chain which may leak the netlist to those fab-based adversaries.
Therefore, following a \emph{strong
threat model}~\cite{imeson13,li18} (reviewed next),
we leverage 3D integration to hinder such an advanced HT threat scenario.

\subsection{Strong Threat Model}
\label{sec:strong_model_prior_art}

\begin{figure}[tb]
\centering
\includegraphics[width=.95\columnwidth]{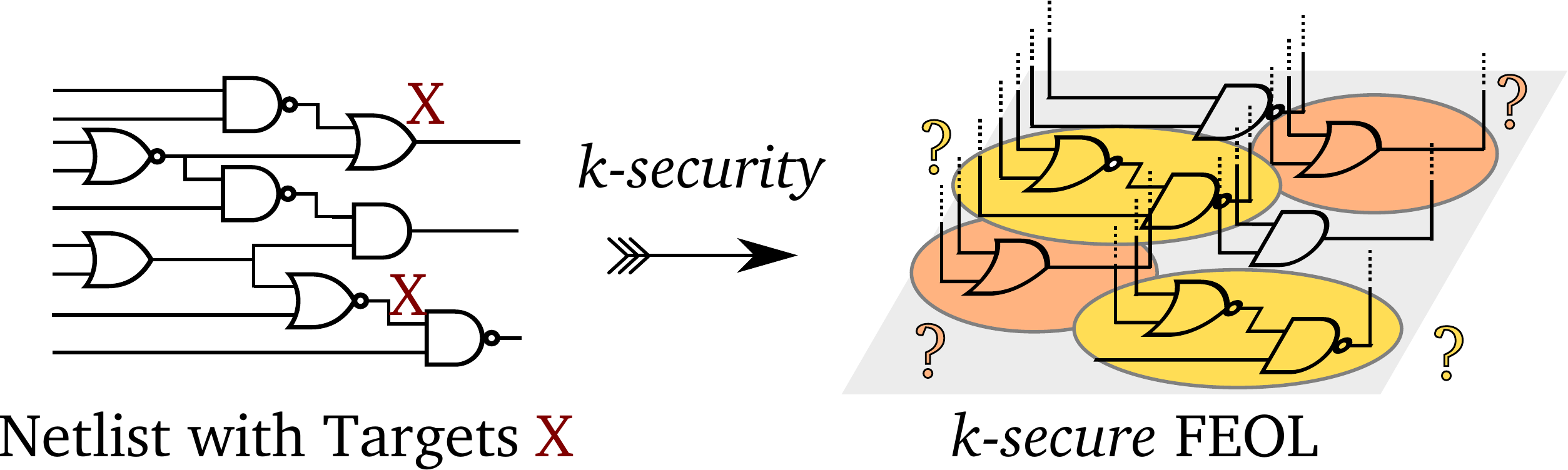}
\caption{\emph{k-security} is to apply split manufacturing and purposefully lift wires to the BEOL (indicated by dashed wires)
such that FEOL-based attackers cannot uniquely identify some or any parts of the netlist available to them. Hence, targeted Trojan insertion becomes difficult. 
Here, with security level two, an attacker has a 50\% chance 
each for correctly inserting Trojans into the targeted OR and NOR-NAND structures.
\label{fig:k-security}
}
\end{figure}

The security guarantee of \emph{k-security}, as proposed by Imeson \emph{et al.}~\cite{imeson13} 
is as follows.
Given a \emph{k-secure} FEOL layout and the complete, final gate-level netlist (just before splitting into FEOL/BEOL),
an attacker has only a chance of $1/k$ for successful HT insertion into a particular location
(or an up to $k$ times higher risk for having $\leq k$ HTs detected by subsequent inspection).
To achieve this, the idea
is to induce $k$ isomorphic structures in the FEOL by carefully lifting wires to the BEOL.
As a result, an attacker cannot uniquely map these $k$ structures to
the target in the netlist, but can only randomly guess
with a probability of $1/k$ (Fig.~\ref{fig:k-security}).
Imeson \emph{et al.}~\cite{imeson13} 
developed a greedy heuristic to select wires to be lifted to the BEOL, and also apply SAT to compute the security level, i.e., the minimal degree of isomorphism for any 
cell type found in the whole FEOL layout.

\subsubsection{Discussion of Prior Art}
Although it provides a formal foundation against HT insertion, \emph{k-security}~\cite{imeson13} has practical shortcomings as follows.
\begin{enumerate}
\item[``-'']
\label{item:shortcoming_synthesis}
The security level hinges on the underlying composition of gates which, in turn, is dictated by logical synthesis.
The fewer the instances of some gate type in the FEOL, the lower the overall security level.
That is because \emph{k-security} is defined as the minimum security level across all gate types.
The technology library plays a key role here as well---the ``richer'' the library in terms of gate types,
the lower the average counts for some individual types, and the lower the overall security level.
We report on this correlation in Table~\ref{tab:min_max_ksecurity}.
\emph{In short, being agnostic to synthesis, \cite{imeson13} is at the mercy of the design tools.}
\item[``-'']
The reported layout cost is considerable. 
For the \emph{ISCAS-85} benchmark \emph{c432}, for 
security level 8, the overheads are
already 61\%, 82\%, and 207\% for PPA, respectively. 
Hence, applying \emph{k-security} on large designs can become 
prohibitive.
\item[``-'']
Computing the security level is \emph{NP-complete} and related to the problem of subgraph isomorphism.
Imeson \emph{et al.}~\cite{imeson13}
leverage SAT solvers
and circuit partitioning
to limit the computational cost.
Still, we note that
the authors did not investigate large-scale benchmarks 
with hundreds of thousands
of gates. Also, while circuit partitioning is a common practice, here 
it hinders
to explore the security for the whole netlist holistically, as sub-circuits can only be secured individually after partitioning.
\item[``-'']
Imeson \emph{et al.}~\cite{imeson13} delegate only wires but no gates to the trusted 2.5D facility.
This approach is the same as with classical SM in 2D ICs, hence the complete potential of 3D integration is not utilized.
\end{enumerate}

\begin{table*}[tb]
\centering
\scriptsize
\setlength{\tabcolsep}{0.5mm}
\caption{Gate composition dictating \emph{k-security} level for various benchmarks under different libraries as in~\cite{imeson13}.
All gates have driving strength X1. 
The maximal security levels equal the minimum count across the libraries' gate types (marked in boldface); the actual level may be lower.
}
\label{tab:min_max_ksecurity}
\smallerspacecaption
\begin{tabular}{|*{24}{c|}}
\hline
\multirow{2}{*}{\textbf{Benchmark}}
& \multicolumn{3}{|c|}{\textbf{lib-3}} 
& \multicolumn{4}{|c|}{\textbf{lib-4}} 
& \multicolumn{5}{|c|}{\textbf{lib-5}}
& \multicolumn{7}{|c|}{\textbf{lib-7}} \\
\cline{2-20}
& \textbf{NOR2} & \textbf{NAND2} & \textbf{INV} 
& \textbf{NOR2} & \textbf{NAND3} & \textbf{NAND2} & \textbf{INV} 
& \textbf{NOR3} & \textbf{NOR2} & \textbf{NAND3} & \textbf{NAND2} & \textbf{INV}  
& \textbf{NOR4} & \textbf{NOR3} & \textbf{NOR2} & \textbf{NAND4}  
& \textbf{NAND3} & \textbf{NAND2} & \textbf{INV} \\ \hline 
\hline

c432 & 61 & 69 & \textbf{30}  
& 43 & \textbf{14} & 65 & 24  
& 15 & 37 & \textbf{9} & 50 & 29 
& \textbf{5} & 9 & 45 & \textbf{5} & \textbf{5} & 37 & 31 
\\ \hline

c7552 & \textbf{461} & 967 & 538  
& 244 & \textbf{158} & 986 & 586  
& \textbf{36} & 227 & 126 & 985 & 527 
& \textbf{2} & 7 & 230 & 59 & 179 & 940 & 523
\\ \hline

b14 & \textbf{806} & 3,094 & 1,282  
& 450 & \textbf{422} & 2,503 & 1,092  
& \textbf{86} & 370 & 362 & 2,587 & 1,151 
& \textbf{7} & 17 & 259 & 88 & 303 & 2,598 & 1,122 
\\ \hline

b17\_1 & 7,097 & 12,909 & \textbf{4,206}
& 5,549 & \textbf{3,179} & 8,500 & 4,195  
& \textbf{1,440} & 4,806 & 2,462 & 7,759 & 4,454 
& \textbf{281} & 585 & 3,733 & 1,070 & 1,394 & 8,533 & 4,339 
\\ \hline

b22 & \textbf{2,405} & 10,818 & 4,618  
& \textbf{1,433} & 1,458 & 9,129 & 4,093  
& \textbf{223} & 1,398 & 1,347 & 8,599 & 3,953 
& \textbf{71} & 92 & 1,183 & 361 & 1,033 & 8,823 & 3,825
\\ \hline

\end{tabular}
\end{table*}

Li \emph{et al.}~\cite{li18} recently proposed an advancement for \emph{k-security}, with the following contributions (labeled as
``+'').
Despite the advances proposed in~\cite{li18}, there are also some limitations (labeled as ``-'').

\begin{enumerate}

\item[``+'']
Additional dummy gates (and dummy wires) help to raise the security level as required 
for layouts initially holding only a few instances of some gate type.
To model the insertion of dummy gates and simultaneous lifting of wires, Li \emph{et al.}~\cite{li18} consider the concept of \emph{spanning subgraph isomorphism}.
\item[``+'']
To limit computational cost, they propose a mixed-integer linear program (MILP)-based framework.
\item[``+'']
To limit layout cost, they protect only selected vulnerable gates in the layout, not the whole design.
They further impose a close but uniform arrangement of isomorphic instances, to limit placement disturbances without leaking
layout-level hints.

\item[``-'']
As with Imeson \emph{et al.}~\cite{imeson13}, Li \emph{et al.}~\cite{li18} apply partitioning on larger netlists, effectively hindering to secure industrial designs in their entirety.
Also, Li \emph{et al.}~\cite{li18} do not investigate any large-scale designs.

\item[``-'']
While Li \emph{et al.}~\cite{li18} may raise the security level using dummy gates (and dummy wires),
these gates can impose significant area overheads (and timing overheads, due to routing congestion when having to lift all related wires).
We report on the PPA impact for adding dummy gates/wires in Table~\ref{tab:meng_li_study}.
For the \emph{ITC-99} benchmark \emph{b17\_1}, for security level 20, there are up to 1.2$\times$ more gates and 5.2$\times$ more wires when
compared to the unprotected designs.\footnote{In~\cite{imeson13} it is only mandated that each vulnerable gate shall have $k-1$ isomorphic
instances, whereas Li \emph{et al.}~\cite{li18} further require that none of these instances are vulnerable themselves which, arguably, provides
a more stringent security notion. 
Dummy gates/wires are necessary for this purpose, to compensate whenever some of the
isomorphic instances are vulnerable themselves.  
In this work, we follow the notion of~\cite{imeson13} while our methods
can be easily tailored toward~\cite{li18} as well.}
 \item[``-'']
Li \emph{et al.} do not utilize the potential of 3D integration.
\end{enumerate}

\begin{table*}[tb]
\centering
\scriptsize
\caption{Study on~\cite{li18}, netlists provided as courtesy
by the authors of~\cite{li18}.
Setup lib-3 is as in Table~\ref{tab:min_max_ksecurity}, and lib-8 contains NAND2, NOR2, AND2, OR2, XOR2, XNOR2, INV, and BUF, all
in X1 strength.
Left:~number of gates and wires for original, unprotected layouts.
Right:~number of additional dummy gates (D.\ Gates) and lifted wires (L.\ Wires) for security levels 10 (S10) and 20 (S20) when 10\%
of original gates are considered vulnerable.
}
\label{tab:meng_li_study}
\smallerspacecaption
\begin{tabular}{|c||c|c|c|c||c|c|c|c|c|c|c|c|}
\hline
\multirow{2}{*}{\textbf{Benchmark}} 
& \multicolumn{2}{|c|}{\textbf{lib-3 (Original)}}  
& \multicolumn{2}{|c||}{\textbf{lib-8 (Original)}} 
& \multicolumn{2}{|c|}{\textbf{lib-3 (S10)}} 
& \multicolumn{2}{|c|}{\textbf{lib-8 (S10)}}
& \multicolumn{2}{|c|}{\textbf{lib-3 (S20)}}
& \multicolumn{2}{|c|}{\textbf{lib-8 (S20)}} \\
\cline{2-13}
 &
 \textbf{Gates} &
 \textbf{Wires} &
 \textbf{Gates} &
 \textbf{Wires} &
 \textbf{D. Gates} &
 \textbf{L. Wires} &
 \textbf{D. Gates} &
 \textbf{L. Wires} &
 \textbf{D. Gates} &
 \textbf{L. Wires} &
 \textbf{D. Gates} &
 \textbf{L. Wires} \\
 \hline \hline

b14 &
5,182 & 5,457 &
4,125 & 4,400 &
3,620 & 12,600  &
4,706 & 14,892 &
7,819  & 25,012 &
9,885 & 29,737 \\ \hline

b15\_1 &
7,722 & 8,207 & 
6,978 & 7,463 &
6,422 & 21,480  &
7,185 & 22,205 &
13,708  & 43,124 &
15,246 & 44,306 \\ \hline

b17\_1 &
24,212 & 25,664 & 
21,500 & 22,952 &
19,988 & 66,271  &
22,648 & 71,341 &
42,216  & 132,153 &
47,867 & 142,487 \\ \hline

b20 &
11,810 & 12,332 & 
10,686 & 9,226 &
8,340 & 27,818  &
9,748 & 33,661 &
17,808  & 55,602  &
22,405 & 67,044 \\ \hline

b22 &
17,841 & 18,608 & 
14,457 & 15,224 &
12,923 & 44,153  &
16,333  & 50,858 &
27,299  & 87,915 &
34,284 & 101,352 \\ \hline

\end{tabular}
\\ \scriptsize
\end{table*}

In short, Imeson \emph{et al.}~\cite{imeson13} and Li \emph{et al.}~\cite{li18} provide a solid formal foundation to protect
against targeted HT insertion, but there are practical limitations to both schemes.

\subsubsection{Our Contributions}
Here we tackle all the outlined shortcomings of~\cite{imeson13,li18}.
There are two key pillars for our work: (i) a security-driven synthesis strategy and
(ii) an end-to-end CAD flow for preventing HTs in 3D ICs.
The motivation for a security-driven synthesis strategy is that~\cite{imeson13,li18} 
implement their protection on top of a given 
netlist, solely as an afterthought.
In contrast, by delegating the construction of isomorphic structures to the synthesis stage,
we effectively render the protection against HTs a design-time priority.
Besides, our end-to-end CAD flow for
3D ICs, extended from the earlier part of this work,
effectively raises the notion of \emph{k-security} toward practical application for preventing HT insertion in large-scale designs.

\subsection{Security-Driven Synthesis Stage}
\label{sec:synth_strategy}

\begin{figure}[tb]
\centering
\includegraphics[width=.9\columnwidth]{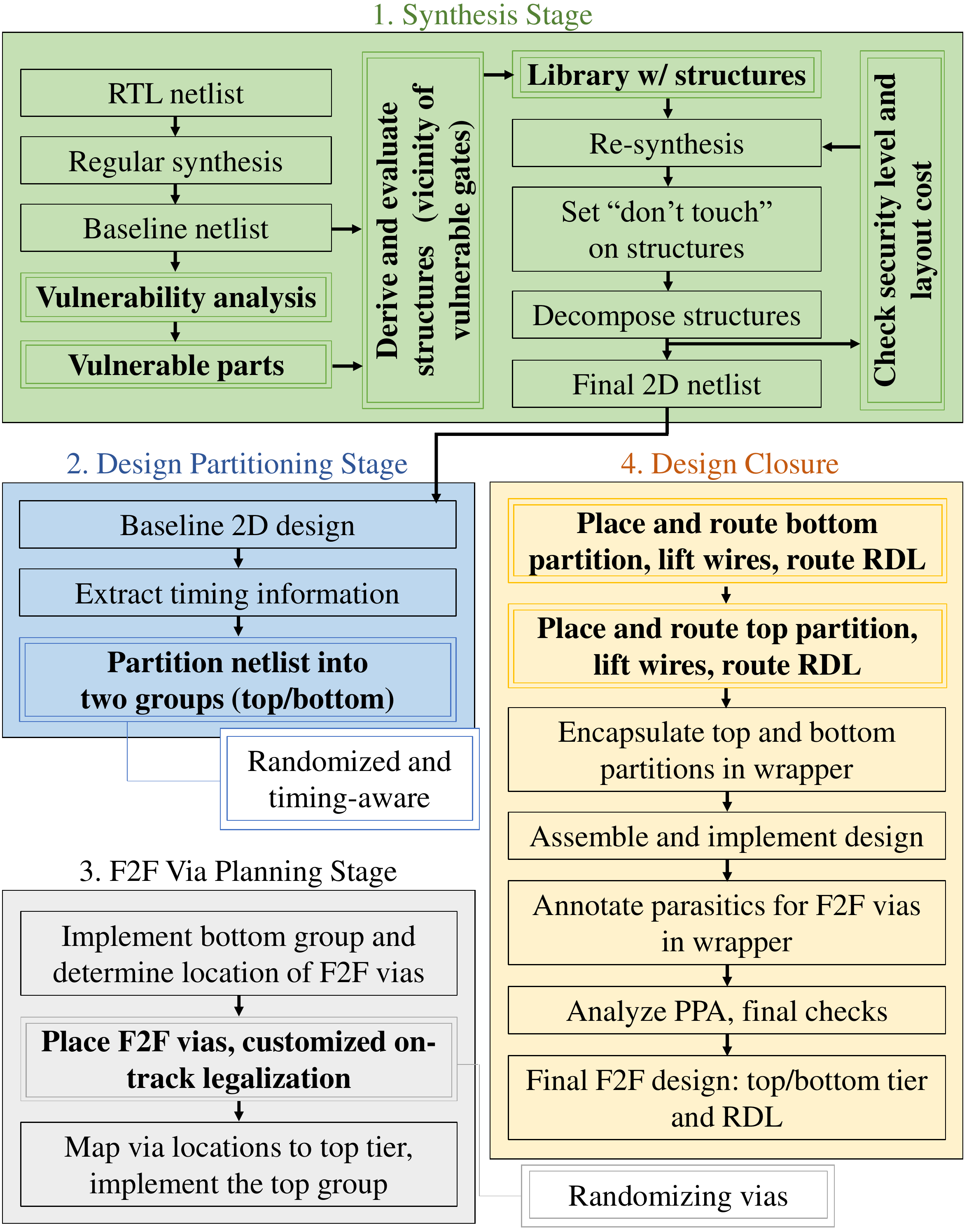}
\caption{Our CAD flow for F2F 3D ICs, with
security-driven steps for preventing HT insertion emphasized in bold.
\label{fig:3D_flow_HT}
}
\end{figure}

The essence of our security-driven synthesis strategy
is as follows.
For any netlist, we 
state that the designer
can identify the structures vulnerable to HT insertion (e.g., by vulnerability analysis~\cite{li18,salmani16}) and wants to
protect them accordingly.
In agreement with \emph{k-security}, the designer then intends to induce many isomorphic instances of those structures in the FEOL. 
To take control of the layout cost, but also to advance scalability and the attainable level of security, we
delegate this step of inducing isomorphic instances
to the synthesis stage.
This way, our approach can be considered as ``secure by construction.''\footnote{Although we
do modify the netlist, we still assume---in agreement with~\cite{imeson13,li18}---that the attacker holds the final gate-level
netlist, truthfully representing our security-driven synthesis stage.
As a result, we do \emph{not} imply security through obscurity.}
We provide more details for the synthesis stage in Fig.~\ref{fig:3D_flow_HT} 
and below.

\begin{figure*}[tb]
\centering
\includegraphics[width=.9\textwidth]{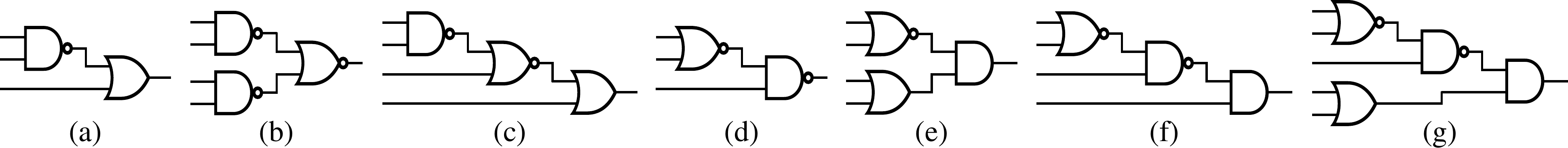}
\caption{The seven structures considered in this work without loss of generality. They are constructed based on the vulnerability analysis
	of~\cite{li18}.
\label{fig:structures}
}
\end{figure*}

Based on some vulnerability analysis of choice~\cite{li18,salmani16}, the designer first identifies the vulnerable gates/structures.  In this study, we leverage~\cite{li18}
to identify various structures which are covering the vulnerable gates as well as some surrounding gates
(Fig.~\ref{fig:structures}). It is understood that the designer can investigate as many structures as desired 
regarding (i)~layout cost,
(ii)~the potential for
inducing isomorphic instances, and (iii)~the coverage of vulnerable gates.
In fact, we explored in total 18 structures; the ones illustrated
in Fig.~\ref{fig:structures} are the most promising ones for our empirical study.\footnote{See Sec.~\ref{sec:experiments_synth} for	the layout cost of all the 18 structures investigated.}

Next, these structures are prepared for synthesis, i.e., they are defined as \emph{custom cells}.
Typically, one has to conduct \emph{library characterization} when creating custom cells,
but here we can refrain from this effort.
That is because
we \emph{decompose} the structures again later on,
i.e., we transform them back into their corresponding arrangement of simple two-input cells.
It is essential, however, to track and preserve all gates related to the decomposed structures, by setting them as ``don't touch,'' such
that the design tools cannot interfere with these gates. We also keep track of the input/output wires of the structures, which have to be
lifted later on to the BEOL to achieve \emph{k-security}.
Now, instead of library characterization, we leverage the characteristics of simple cells available in the library (e.g., NAND with the same
number of inputs as the structure), but we adapt the Boolean functionality as needed.
We note that
utilizing the characteristics of simple cells
saves not only effort but also ``tricks'' the synthesis tool, at least
to some degree, into using more instances for those structures
(see also Sec.~\ref{sec:experiments_synth},
 Table~\ref{tab:structure_count}).
That is presumably because the structures provide some complex Boolean functionality with little, indeed over-optimistic layout cost.
With our iterative synthesis approach,
we can thus impose more instances of the various structures as
needed, while also correctly gauging the anticipated layout cost (by decomposing the structures).

\begin{table*}[ht]
\centering
\scriptsize
\setlength{\tabcolsep}{0.42em}
\caption{Isomorphic instances of structures as in Fig.~\ref{fig:structures}, netlist coverage, and final security level \emph{k} for the iterative re-synthesis.
}
\smallerspacecaption
\label{tab:structure_count}
\begin{tabular}{|c|c|c|c|c|c|c|c|c|c|c|c|c|c|c|c|c|c|c|c|}
\hline
\multirow{2}{*}{\textbf{Benchmark}} 
& \multicolumn{7}{|c|}{\textbf{Iteration 1}} 
& \multicolumn{7}{|c|}{\textbf{Iteration 5}}
& \multicolumn{5}{|c|}{\textbf{Design Coverage, Security Level After Iteration 5}} \\
\cline{2-20}
 &
 \textbf{(a)} &
 \textbf{(b)} &
 \textbf{(c)} &
 \textbf{(d)} &
 \textbf{(e)} &
 \textbf{(f)} &
 \textbf{(g)} &
 \textbf{(a)} &
 \textbf{(b)} &
 \textbf{(c)} &
 \textbf{(d)} &
 \textbf{(e)} &
 \textbf{(f)} &
 \textbf{(g)} &
 \textbf{(a)--(g)} &
 \textbf{Gates Covered} &
 \textbf{Total Gates} &
 \textbf{Coverage} &
 \textbf{\emph{k}} \\
 \hline \hline

b14 &
 89 & 39 & 16 & 79 & 151 & 29 & \textbf{1} &
 137 & 43 & \textbf{30} & 164 & 179 & 65 & 33 &
 651 & 1,780 & 5,061 & 35.17\% & 30
\\ \hline

b15\_1 &
  319 & 106 & \textbf{37} & 93 & 178 & 73 & 117 &
  365 & 112 & \textbf{46} & 139 & 206 & 80 & 138 &
 1,086 & 3,018 & 7,639 & 39.51\% & 46
\\ \hline

b17\_1 &
  1,140 & 309 & \textbf{127} & 318 & 485 & 161 & 388 &
  1,239 & 333 & \textbf{164} & 424 & 532 & 207 & 442 &
   3,341 & 9,173 & 23,912 & 38.36\% & 164
\\ \hline

b18 &
  2,170 & 778 & \textbf{411} & 612 & 1,097 & 487 & 952  &
  2,583 & 870 & \textbf{576} & 1,046 & 1,243 & 712 & 1,156 &
  8,186 & 23,373 & 67,775 & 34.49\% & 576
\\ \hline

b19 &
  5,011 & 1,558 & \textbf{939} & 1,320 & 2,422 & 1,174 & 1,669  &
  5,809 & 1,745 & \textbf{1,221} & 2,420 & 2,832 & 1,762 & 2,073 &
  17,862 & 50,413 & 140,176 & 35.96\% & 1,221
\\ \hline

b20 &
  164 & 85 & \textbf{27} & 206 & 208 & 86 & 0 &
  308 & 99 & \textbf{70} & 425 & 277 & 180 & 72 &
  1,431 & 3,882 & 11,116 & 34.92\% & 70
\\ \hline

b22 &
 249 & 85 & 47 & 345 & 340 & 149 & \textbf{7} &
  453 & \textbf{108} & 114 & 727 & 454 & 325 & 124 &
  2,305 & 6,298 & 16,941 & 37.18\% & 108
\\ \hline

\end{tabular}
\end{table*}

Once the synthesis iterations are completed, which is upon the designer to decide and should be based on the security level and/or synthesis-level PPA cost,
the vulnerability of the final netlist is to be re-evaluated.
The final security level hinges on how many vulnerable gates are covered by the isomorphic instances of all structures.

\subsection{End-to-End CAD Flow}
\label{sec:HT_flow}

As indicated, our 3D IC CAD flow
for preventing targeted HT insertion
is
extended from the earlier part of this work.
However, there are some differences as follows.

\begin{enumerate}
\item
Since the threat of HT insertion applies exclusively to manufacturing time,
we do not require LC/obfuscation for the RDL.
However, we require randomization of the routing paths within the RDL. 
\item \emph{k-security} is applied initially, to the whole netlist at once, using our security-driven synthesis strategy.
\item
Our flow comprises techniques similar to those required for regular SM. More specifically, we
leverage our customized \emph{lifting cells}~\cite{patnaik18_SM_ASPDAC} to lift wires to the RDL as dictated by \emph{k-security}.
\end{enumerate}

Our flow enables the concerned designer to tackle both layout cost and resilience against HT insertion.
To do so, we integrate our security-driven
synthesis strategy as a key stage for the overall flow
(Fig.~\ref{fig:3D_flow_HT}).
Further steps in our flow are partitioning, planning of F2F vias, and placement and SM-aware routing.
We provide some details below.

\textbf{Design partitioning:}
After conducting the security-driven synthesis strategy,
	we place and route the 2D design.
The resulting netlist is partitioned into two groups, representing the top and bottom tiers of the F2F 3D IC.
We define I/O ports for wires crossing the two tiers after partitioning; these ports become F2F vias later on.

As for the impact of 3D partitioning on \emph{k-security}, it is important to note the following.
First, we ensure that all the gates of any decomposed isomorphic instance stay together in one tier.
Second, since we apply the synthesis strategy on the complete netlist, i.e., before partitioning, there is no inherent
limitation for isomorphic instances.
Third, partitioning itself may, however, still impact the final security level. That is because once an attacker can retake the same
partitioning steps as the designer, the attacker can also infer which \emph{subset} of isomorphic instances goes into which tier.
It is easy to see that fully random partitioning would render this attacker's benefit void, but we found
that this can impose considerable layout cost.
Based on Sec.~\ref{sec:design_partitioning}, we therefore propose a customized timing-driven and security-aware partitioning technique as follows.

First, we obtain the timing reports for the 2D baseline layout.
Then, each critical timing path without any isomorphic instances is kept within one tier.
Other paths,
i.e., paths with some isomorphic instances or non-critical paths,
are randomly partitioned across the two tiers.
With this partitioning technique, the attacker cannot understand which isomorphic instances in the bottom/top tier relate to which in the
netlist---the definition of the security level as in~\cite{imeson13} is maintained.

\textbf{Planning of F2F interconnects:}
These steps are primarily the same as in Sec.~\ref{sec:planning_F2F}, in particular, the randomization of F2F ports and the custom on-track
legalization, but here we can refrain from implementing obfuscated switchboxes.

\textbf{Wire lifting and design closure:}
Next, both tiers are placed and routed separately.
In the absence of the obfuscated switchboxes leveraged for routing the RDL in Sec.~\ref{sec:closure},
here we re-tailor our custom lifting cells~\cite{patnaik18_SM_ASPDAC}
to enable wire-lifting as required for \emph{k-security}.
That is, while routing the top/bottom tier, we route the regular metal layers, lift wires to the RDL
with the help of the lifting cells, and route the RDL.
For design closure, the top and bottom tiers are wrapped into one netlist.  Again, we purposefully
\emph{do not engage in any optimization across tiers}, to \emph{maintain anonymized layouts}.
As in Sec~\ref{sec:closure}, we derive the SPEF
from the wrapper netlist to capture the RC parasitics of F2F vias, and we evaluate the final layout cost.

\section{Results for Trojan Prevention}
\label{sec:HT_results}

The setup is the same as in Sec.~\ref{sec:IP_results}, except that two M8 layers are used for the RDL.
That is also because without the need for obfuscated switchboxes, the RDL is less complex.

\subsection{Analysis of the Security-Driven Synthesis Stage}
\label{sec:experiments_synth}

\begin{figure}[tb]
\centering
\includegraphics[width=.98\columnwidth]{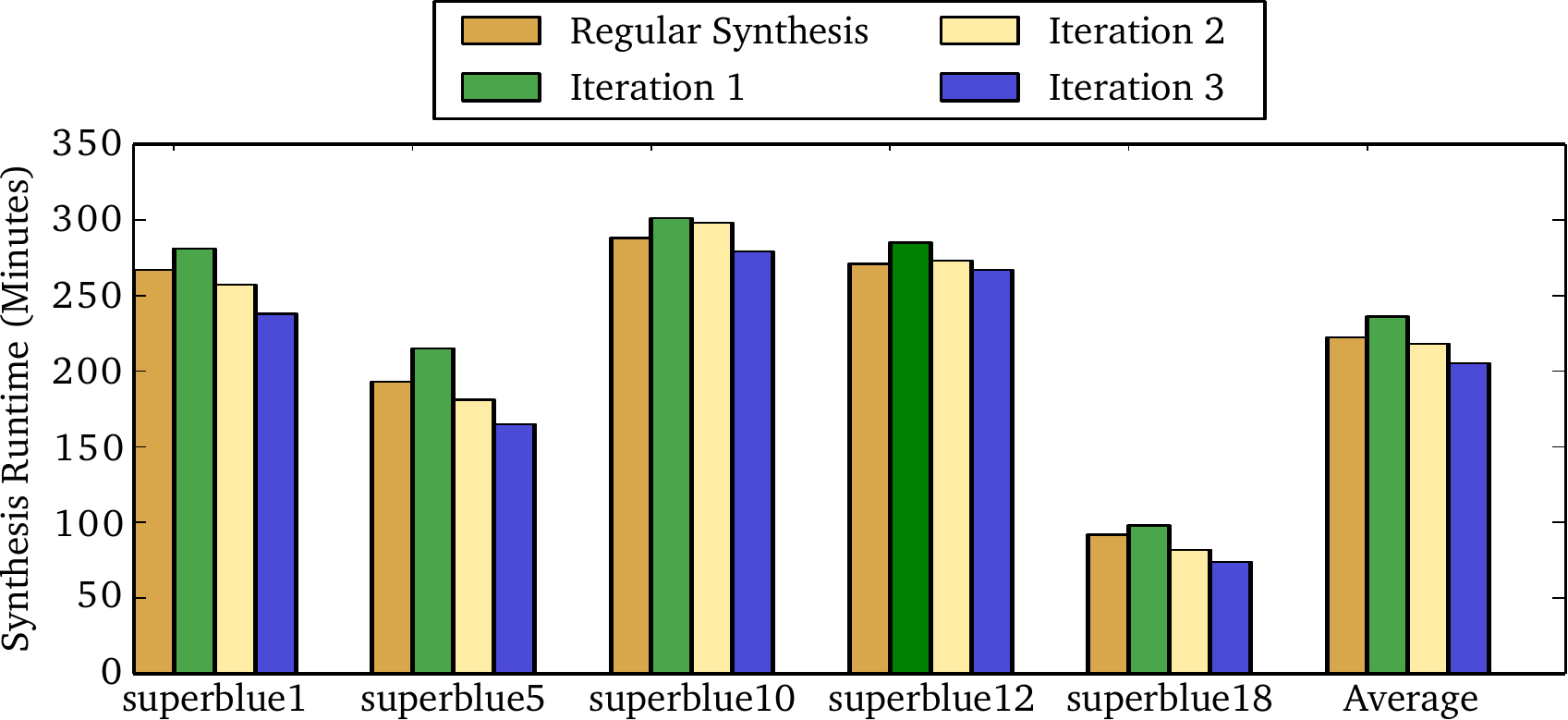}
\caption{Runtime comparison for regular synthesis and our security-driven synthesis strategy for large-scale \emph{IBM superblue}
	benchmarks.
\label{fig:IBM_syn_runtimes}
}
\end{figure}

During the iterative, security-driven synthesis stage, note that we fix/preserve all gates of any isomorphic structure.
We observe that doing so helps to guide the logical synthesis toward the remaining parts of the netlist not yet covered by some structures;
we can increase the instance counts within a reasonable runtime.
Since the synthesis iterations require only a few minutes for all the commonly considered benchmarks (e.g., 25--45 minutes even for \emph{ITC-99}),
we additionally explore the large-scale \emph{IBM superblue} benchmarks for
scalability of our synthesis strategy (Fig.~\ref{fig:IBM_syn_runtimes}).
Here we observe that our strategy still impose only little runtime cost, about 6.3\% on average for the first iteration,
and runtimes for successive iterations are further reducing.

In Table~\ref{tab:structure_count}, we report on isomorphic instances and their coverage for various benchmarks, based on
10\% of all gates being identified as vulnerable ones~\cite{li18}.
We find that large parts of the netlists can be covered by isomorphic instances already after a few synthesis iterations, namely 36.5\%, on
average, after five iterations. 
This coverage provides strong
protection beyond the 10\% of gates targeted.
For the large-scale \emph{IBM superblue} benchmarks
(not illustrated in Table~\ref{tab:structure_count})
the coverage is even higher, at 59\%, on average.
Although we seek to cover all the gates which have been initially identified as vulnerable ones,
some of those gates may not be covered in the final netlist, due to the iterative re-synthesis runs.
We found that this effect is acceptable;
of the 10\% identified gates,
there are 9\% covered on average.
Besides, we note that few of the critical paths contain isomorphic instances to begin with.
That is because critical paths rarely
contain vulnerable gates~\cite{li18,salmani16}; related Trojans would be easy to detect by delay testing.

As for security levels following the definition in~\cite{imeson13},
we can achieve significant levels already after the fifth iteration, ranging
from 30 for \emph{ITC-99} benchmarks \emph{b14} up to 1,221 for \emph{b19}.
Based on the counts of individual instances, one may also revisit
the synthesis after ruling out some structures which tend to limit the security levels, e.g., structure (c) for
Table~\ref{tab:structure_count}.
Hence, our strategy provides margin to the designer for both cases when lower levels are sufficient or higher levels are desired.

In Table~\ref{tab:PPA_comparison_iterative_resynth}, we report on the impact of some synthesis iterations and the final 2D layout cost.  For
the latter, wire-lifting toward the BEOL as required by \emph{k-security} is already accounted for.
Overall,
layout costs are acceptable; that is especially true when qualitatively comparing to~\cite{imeson13}, where already notably smaller
benchmarks induced significant PPA cost.
That is also because we can safely apply buffer insertion to tackle timing degradation, as can be seen below.

\begin{table*}[tb]
\centering
\scriptsize
\setlength{\tabcolsep}{0.2em}
\caption{Gate counts and associated layout cost. Final layout-level cost account for 
	wire-lifting (WL) in 2D. All cost are in \% and with respect to the 2D baseline.
	A refers to area, P to power, and D to delay.
}
\smallerspacecaption
\label{tab:PPA_comparison_iterative_resynth}
\begin{tabular}{|c|c|c|c|c|c|c|c|c|c|c|c|c|c|c|c|c|c|c|c|c||c|c|c|}
\hline
\multirow{2}{*}{\textbf{Benchmark}} 
& \multicolumn{4}{|c|}{\textbf{Iteration 1}}
& \multicolumn{4}{|c|}{\textbf{Iteration 2}} 
& \multicolumn{4}{|c|}{\textbf{Iteration 3}} 
& \multicolumn{4}{|c|}{\textbf{Iteration 4}} 
& \multicolumn{4}{|c||}{\textbf{Iteration 5}}
& \multicolumn{3}{|c|}{\textbf{Final (with WL)}} \\
\cline{2-24}
 &
 \textbf{Gates} &
 \textbf{A} &
 \textbf{P} &
 \textbf{D} &
 \textbf{Gates} &
 \textbf{A} &
 \textbf{P} &
 \textbf{D} &
 \textbf{Gates} &
 \textbf{A} &
 \textbf{P} &
 \textbf{D} &
 \textbf{Gates} &
 \textbf{A} &
 \textbf{P} &
 \textbf{D} &
 \textbf{Gates} &
 \textbf{A} &
 \textbf{P} &
 \textbf{D} &
 \textbf{A} &
 \textbf{P} &
 \textbf{D} \\
 \hline \hline

b14 &
4,471 & 8.66 & 13.65 & 5.03 & 
4,683 & 12.78 & 19.11 & 5.10 & 
4,938 & 18.6 & 26.6 & 5.4 &
4,967 & 19.72 & 27.7 & 5.14 &
5,061 & 21.62 & 28.64 & 8.75
& 21.62 & 62.36 & 55.65
\\ \hline

b15\_1 &
7,535 & 11.23 & 8.17 & 7.7 & 
7,622 & 12.67 & 10.95 & 7.58 & 
7,640 & 13.33 & 12.87 & 7.6 &
7,658 & 14.39 & 13.12 & 7.65 &
7,722  & 16.25 & 15.23 & 7.84
& 16.25 & 43.24 & 61.65
\\ \hline

b17\_1 &
23,491 & 13.21 & 11.67 & 7.55 & 
23,876 & 14.94 & 14.07 & 8.05  & 
23,960 & 15.00 & 15.57 & 7.54 & 
24,250 & 15.58 & 19.15 & 9.03 & 
25,112 & 18.95 & 23.18 & 15.5  
& 93.95 & 58.96 & 63.54
\\ \hline

b18 &
64,417 & 6.51 & -10.87 & 17.64 & 
66,222 & 9.32 & 3.47 & 17.57  & 
67,134 & 10.01 & 4.04 & 14.74 & 
67,671 & 11.68 & 7.06 & 16.08 & 
67,775 & 12.66 & 8.14 & 15.67  
& 112.66 & 41.48 & 75.18
\\ \hline

b19 &
134,634 & 0.47 & -29.9 & 24.27 & 
137,478 & 2.73 & -15.36 & 21.75  & 
139,032 & 3.02 & -13.94 & 20.61 & 
140,016 & 3.71 & -12.81 & 21.80 & 
140,176 & 4.59 & -13.3 & 21.37  
& 154.59 & 51.17 & 88.87
\\ \hline

b20 &
9,716 & 4.94 & 4.32 & 3.33 & 
10,268 & 9.61 & 11.68 & 3.31  & 
10,835 & 15.13 & 17.44 & 3.66 & 
10,981 & 16.61 & 16.25 & 3.47 & 
11,116 & 18.0 & 18.81 & 7.0  
& 18.0 & 54.23 & 63.32
\\ \hline

b22 &
15,058 & 4.93 & 7.88 & 6.23 & 
16,015 & 10.13 & 13.29 & 7.13  & 
16,613 & 13.92 & 18.21 & 7.98 & 
16,941 & 15.96 & 20.17 & 10.6  &
16,963 & 16.14 & 21.72 & 10.72
& 16.14 & 54.44 & 66.61
\\ \hline

\end{tabular}
\end{table*}

\begin{figure*}[tb]
\centering
\includegraphics[width=2\columnwidth]{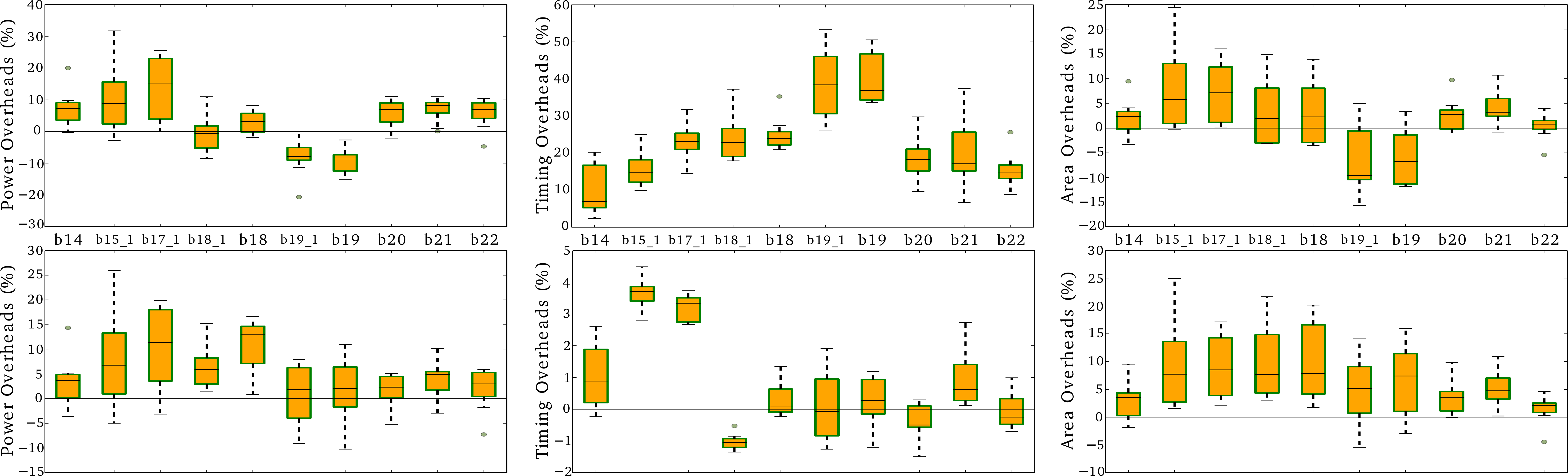}
\caption{
Distribution of layout cost for all 18 different structures we explored on various benchmarks.
Top: without any optimization; bottom: with buffer insertion applied (only outside of the structures).
Note that the same benchmarks are applied for the top and bottom plots; benchmark labels are accordingly placed between those plots.
Also note the different scales for each plot.
Each boxplot represents ten runs.
\label{fig:PPA_structures_boxplot}
}
\end{figure*}

As foreclosed, we also provide the distribution of layout cost for all 18 different structures we explored in
Fig.~\ref{fig:PPA_structures_boxplot}.
There we contrast the re-synthesized netlists without any optimization to those after buffer insertion.
Since all isomorphic
instances are preserved, buffer insertion cannot interfere with those structures and \emph{security is not undermined}.
This simple technique helps to avoid large cost while ``tricking'' synthesis into using our custom cells; it forms the baseline for
the remaining, 3D-centric experiments in Sec.~\ref{sec:analysis_final_HT}

Finally, since \emph{ours is the first to consider large-scale benchmarks} for \emph{k-security}, a direct comparison
with prior works~\cite{imeson13,li18} is impractical.
For a qualitative comparison, our work allows for superior 
security levels, induces little layout cost, and is scalable, all by means of synthesis.

\subsection{3D IC Layout Cost and Security Analysis}
\label{sec:analysis_final_HT}

\begin{figure}[tb]
\centering
\includegraphics[width=\columnwidth]{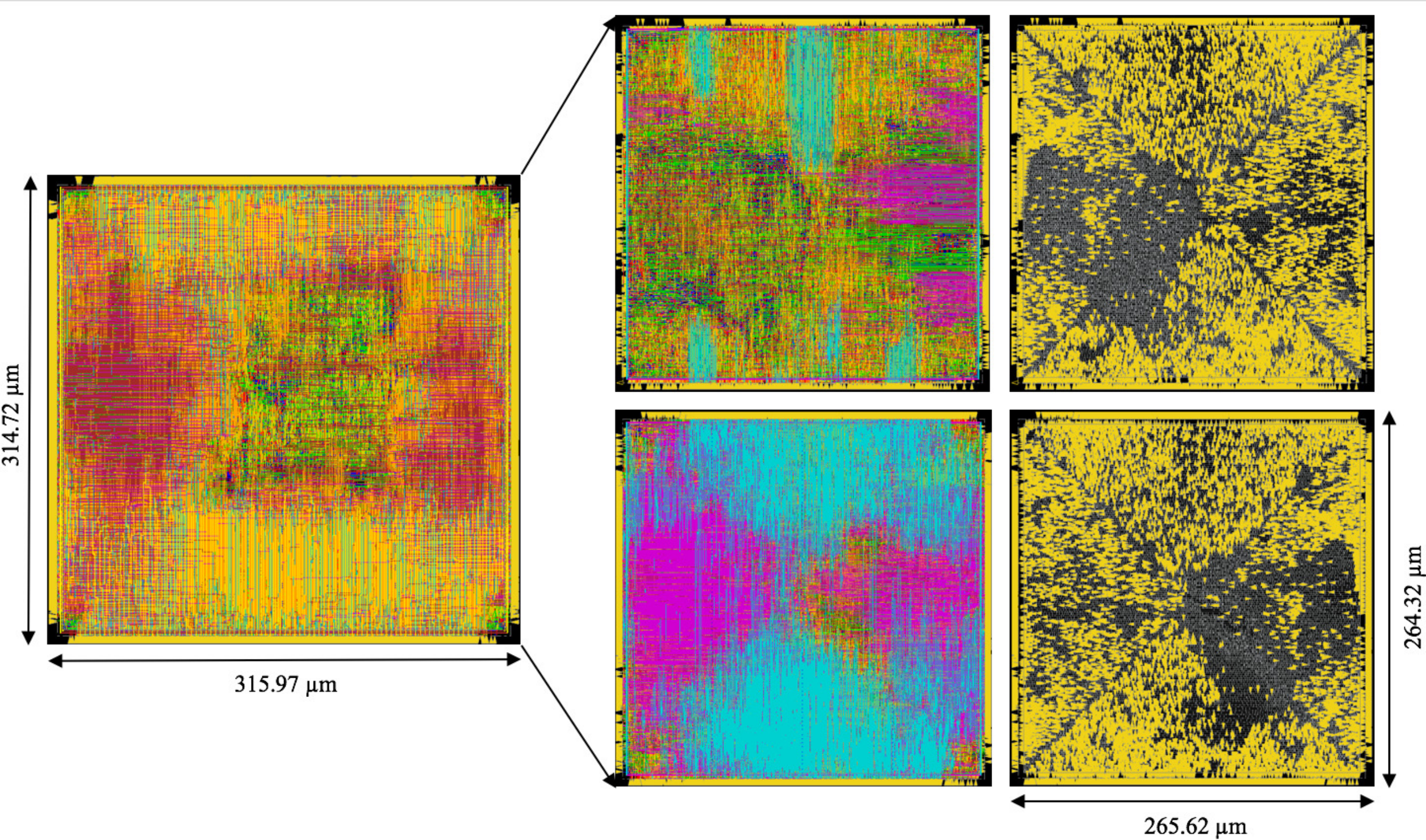}
\caption{2D layout (left), bottom/top tier of the F2F 3D IC (right), both for benchmark b18.
	Yellow dots (very right) illustrate the F2F vias.
\label{fig:HT_die_images}
}
\end{figure}

In Table~\ref{tab:HT_results}, we report on the final layout cost for the F2F 3D ICs.
Overall, costs are better than for the 2D setup
(Table~\ref{tab:PPA_comparison_iterative_resynth}), especially for larger benchmarks. This key finding attests our objective to advance
\emph{k-security} for large-scale designs.
In Fig.~\ref{fig:HT_die_images}, we showcase the layouts for \emph{ITC-99} benchmark \emph{b18}.

\begin{table}[tb]
\centering
\scriptsize
\setlength{\tabcolsep}{0.17em}
\caption{Isomorphic instances, security level, layout cost (in \%) with respect to 2D baseline, and SAT attacks (t-o is 100 hours).
}
\label{tab:HT_results}
\smallerspacecaption
\begin{tabular}{|c|c|c|c|c|c|c|c|c|c|}
\hline
\multirow{2}{*}{\textbf{Benchmark}} 
& \multicolumn{2}{|c|}{\textbf{Structures}}
& \textbf{Security Level$^*$}
& \multicolumn{3}{|c|}{\textbf{Overheads}} 
& \textbf{SAT Attack~\cite{subramanyan15}}\\
\cline{2-3}
 &
 \textbf{Bottom} &
 \textbf{Top} &
 \textit{k} &
 \textbf{Area} &
 \textbf{Power} &
 \textbf{Delay} &
 \textbf{Runtime} \\
 \hline \hline

b14 & 16 & 14 & 30 & -40.32 & 30.66 & 37.17 & t-o \\ \hline
b15\_1 & 27 & 19 & 46 & -40.21 & 27.69 & 38.32 & t-o \\ \hline
b17\_1 & 84 & 80 & 164 & -37.92 & 46.56 & 45.73 & t-o \\ \hline
b18 & 200 & 145 & 345 & -29.39 & 22.16 & 51.09 & t-o \\ \hline
b19 & 200 & 200 & 400 & -25.48 & 8.46 & 62.44 & t-o \\ \hline
b20 & 38 & 32 & 70 & -38.32 & 42.46 & 47.86 & t-o \\ \hline
b22 & 58 & 50 & 108 & -34.47 & 45.52 & 46.92 & t-o \\ \hline
\end{tabular}
\\ \scriptsize
$^*$The level is defined as the sum of the least occuring structures in the bottom and top tier; see also our security-aware partitioning
technique in Sec.~\ref{sec:HT_flow}.
\end{table}

We also report on security levels in Table~\ref{tab:HT_results}. The levels in 3D are the same as in 2D---as explained above, our 3D
partitioning does not undermine security.
In general,
   few if any \emph{proper} attacks on \emph{k-security} are available yet,
and this is because the underlying notion of \emph{k-isomorphism} is formally secure~\cite{cheng10}.
This also implies that otherwise effective attacks will not be applicable. For example,
although we can tackle the missing RDL
connections using a SAT attack as~\cite{subramanyan15}---the netlist available to the adversary can serve as ``virtual
oracle'' here---doing so is not practical.
First, as we show in Table~\ref{tab:HT_results},
	SAT attacks become computationally expensive for large designs.
Second, while a SAT attack may eventually provide a \emph{functionally equivalent} assignment for the RDL connectivity, they cannot
provide the \emph{structurally equivalent} assignment required for attacking \emph{k-security} any better than random guessing
would.

\section{Conclusion and Outlook}
\label{sec:conclusion}

In this work, we demonstrate in detail how 3D integration is a naturally strong match to
combine split manufacturing and camouflaging.
In particular, we promote ``3D splitting'' a design across multiple tiers along with randomization and camouflaging of the
vertical interconnects between the tiers.
By doing so, we propose a modern approach to (i) IP protection and (ii) prevention of targeted Trojan insertion.

Using industrial tools and know-how, we develop security-driven CAD flows for face-to-face (F2F) 3D ICs, allowing us to
tackle these two essential hardware security challenges.
Among other steps, we propose several security-driven partitioning techniques, randomized planning of F2F ports, customized cells for
obfuscation of vertical interconnects using Mg/MgO vias, and a security-driven synthesis strategy. The latter allows us to apply
\emph{k-security} (a formally secure scheme concerning
		Trojan insertion)
for the first time on large-scale designs and also
to protect sensitive design structures of choice readily.

We conduct comprehensive experiments on DRC-clean layouts, using commonly considered benchmarks as well as large-scale, ``real-life''
designs. Strengthened by extensive security analysis, we argue that leveraging 3D integration is highly promising for hardware security.
Finally, we also put forward a practical threat model which accounts for the business practices of present-day design houses.

For future work,
we plan to explore how 3D integration can also provide resilience against physical attacks such
as invasive probing or exploitation of side-channel leakage.

\section*{Acknowledgments}
\label{sec:acknowledgments}

This work was supported in part by NYUAD under REF Grant RE218 and by the NYU/NYUAD joint Center for Cybersecurity (CCS).
We also thank Dr.\ Anja Henning-Knechtel for preparing selected illustrations.

\ifCLASSOPTIONcompsoc

\ifCLASSOPTIONcaptionsoff
  \newpage
\fi

\begin{IEEEbiography}[{\includegraphics[width=1in,height=1.25in,clip,keepaspectratio]
{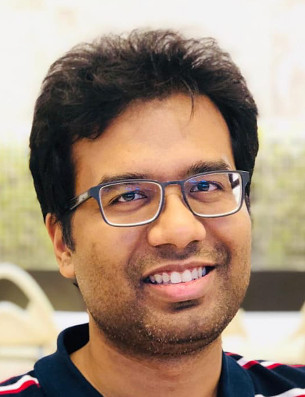}}]{Satwik Patnaik} (M'16)
received B.E.\
in Electronics and Telecommunications from the University of
Pune, Pune, India and
M.Tech.\ in Computer Science and Engineering with a specialization in VLSI Design from Indian Institute of Information Technology and
Management, Gwalior, India. 
He is a Ph.D.\ candidate at the Department of Electrical and Computer Engineering at the 
Tandon School of Engineering with New York University, Brooklyn, 
NY, USA. 
He is a Global Ph.D.\ Fellow with New York University Abu Dhabi, Abu Dhabi, UAE. 
His current research interests 
include Hardware Security, Trust and Reliability issues for CMOS and Emerging Devices 
with particular focus on low-power VLSI Design.
He is a student member of IEEE and ACM.
\end{IEEEbiography}

\begin{IEEEbiography}[{\includegraphics[width=1in,height=1.25in,clip,keepaspectratio]{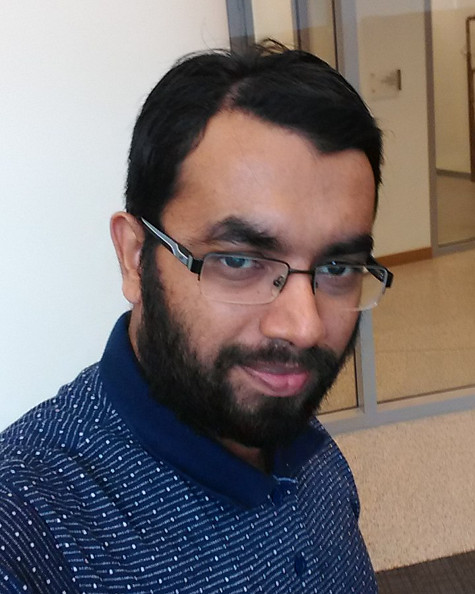}}]{Mohammed Ashraf}
is a 
Senior Physical Design engineer from India. 
He obtained his Bachelor's degree in electronics and 
telecommunication engineering from College of Engineering Trivandrum, Kerala, in 2005. 
He carries an experience of 10 years in the VLSI industry. 
He has worked with various multi-national companies like NVIDIA Graphics, Advanced Micro Devices (AMD), and Wipro Technologies. 
He worked also with Dubai Circuit Design, Dubai Silicon Oasis, UAE.
Mr.\ Ashraf is currently a Research Engineer at Center for Cyber Security (CCS) at New York University Abu Dhabi (NYUAD). 
His work focus on the Physical Design/Implementation of the ARM 
Cortex M0 processor and its four secure variants.
\end{IEEEbiography}

\begin{IEEEbiography}[{\includegraphics[width=1in,height=1.25in,clip,keepaspectratio]{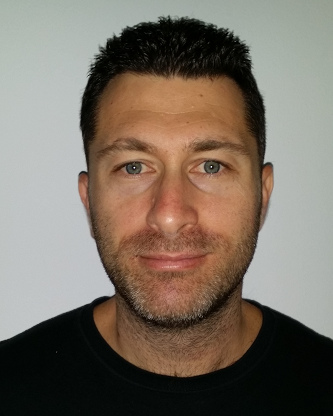}}]{Ozgur Sinanoglu}
is a Professor of Electrical and Computer Engineering at New York University Abu Dhabi. He earned his B.S.\ degrees, one in
Electrical and Electronics Engineering and one in Computer Engineering, both from Bogazici University, Turkey in 1999. He obtained his MS
and PhD in Computer Science and Engineering from University of California San Diego in 2001 and 2004, respectively. He has industry
experience at TI, IBM and Qualcomm, and has been with NYU Abu Dhabi since 2010. During his PhD, he won the IBM PhD fellowship award twice.
He is also the recipient of the best paper awards at IEEE VLSI Test Symposium 2011 and ACM Conference on Computer and Communication Security
2013. 

Prof.\ Sinanoglu's research interests include design-for-test, design-for-security and design-for-trust for VLSI circuits, where he has more
than 180 conference and journal papers, and 20 issued and pending US Patents. Sinanoglu has given more than a dozen tutorials on hardware
security and trust in leading CAD and test conferences, such as DAC, DATE, ITC, VTS, ETS, ICCD, ISQED, etc. He is serving as track/topic
chair or technical program committee member in about 15 conferences, and as (guest) associate editor for IEEE TIFS, IEEE TCAD, ACM JETC,
      IEEE TETC, Elsevier MEJ, JETTA, and IET CDT journals. 

Prof.\ Sinanoglu is the director of the Design-for-Excellence Lab at NYU Abu Dhabi. His recent research in hardware security and trust
is being funded by US National Science Foundation, US Department of Defense, Semiconductor Research Corporation, Intel Corp and
Mubadala Technology.
\end{IEEEbiography}

\begin{IEEEbiography}[{\includegraphics[width=1in,height=1.25in,clip,keepaspectratio]{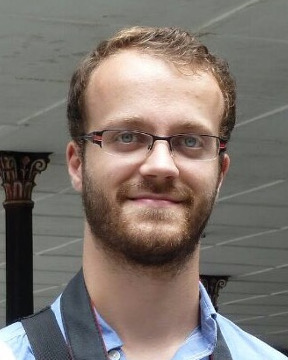}}]{Johann Knechtel}
(M'11)
received the M.Sc.\ in Information Systems Engineering (Dipl.-Ing.) in 2010 and the Ph.D.\ in Computer Engineering
(Dr.-Ing.) in 2014, both from TU Dresden, Germany.  He is currently a Research Associate
at the New York University Abu Dhabi (NYUAD), UAE.  Dr.\ Knechtel was a
Postdoctoral Researcher in 2015--16 at the Masdar Institute of Science and Technology, Abu Dhabi.  From 2010 to 2014, he was
a Scholar with the DFG Graduate School on ``Nano- and Biotechnologies for Packaging of Electronic
Systems'' and the Institute of Electromechanical and Electronic Design, both hosted at the TU Dresden.  In 2012, he was a
Research Assistant with the Dept.\ of Computer Science and Engineering, Chinese University of Hong Kong, China.  In 2010, he
was a Visiting Research Student with the Dept.\ of Electrical Engineering and Computer Science, University of Michigan, USA.
His research interests cover VLSI Physical Design Automation, with particular focus on Emerging Technologies and Hardware Security.
\end{IEEEbiography}

\setcounter{section}{0}
\renewcommand{\thesection}{S.\Roman{section}} 
\renewcommand{\thesubsection}{\thesection.\Alph{subsection}}

\setcounter{figure}{0}
\makeatletter
\makeatletter \renewcommand{\fnum@figure} {\figurename~S\thefigure} 
\makeatother

\setcounter{table}{0}
\makeatletter
\makeatletter \renewcommand{\fnum@table}
{\tablename~S\thetable}
\makeatother

\pagebreak

\section{Study on Different Technology Nodes}
\label{sec:diff_nodes_supp}
Using the \emph{ITC-99} benchmarks, here we gauge the capabilities for heterogeneous 3D SM,
assuming a trusted 180nm foundry and an untrusted 45nm foundry.
More specifically, we leverage
the \emph{OSU} libraries~\cite{OSU_PDK}.
Their libraries hold the same number, type, and strengths of cells; this guarantees a fair comparison since CAD tools cannot leverage
different versions of cells.
\emph{Synopsys DC} was used for synthesis and place and
route was performed using \emph{Cadence Innovus 17.1}; see also Sec.~\ref{sec:IP_setup} for details on the F2F 3D setup.

PPA results for an aggressive timing closure of the 2D baseline setup are given in
Table~S\ref{tab:basic_PPA_two_tech}.
For the heterogeneous F2F 3D setup,
we observe performance degradations as we lift more and more gates to the trusted low-end tier
(Fig.~S\ref{fig:two_tech_delay}).
Also, note from Table~S\ref{tab:basic_PPA_two_tech} that
area (and power) cost is $\approx$12X (and 9X) when contrasting 180nm to 45nm.
To maintain a balanced utilization for both tiers, these findings imply that one should not lift more than $\approx$8\% of the gates to
the low-end tier.
While such small-scale lifting provides a reasonable performance gain, especially from the
perspective of commissioning only the 180nm foundry, it may not be enough to cover all the sensitive design parts.

These findings are in general agreement with those provided in Sec.~\ref{sec:diff_found}.

\begin{figure}[h]
\centering
\includegraphics[width=.8\columnwidth]{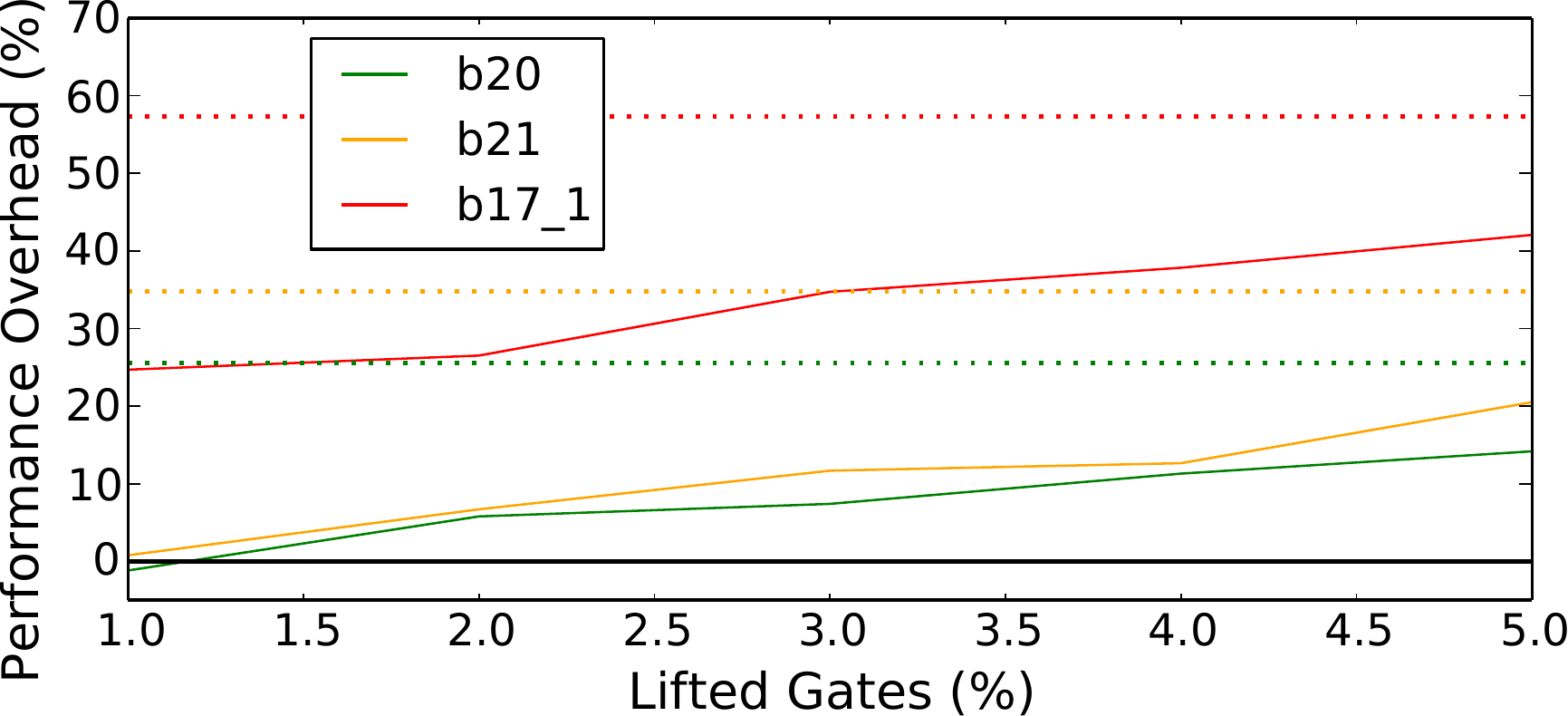}
\caption{Performance degradation when lifting gates in 3D from the 45nm tier to the 180nm tier.
Dotted lines indicate the critical-path delays for the 180nm 2D baseline setup.
See also Table~S\ref{tab:basic_PPA_two_tech} and its footnote.
\label{fig:two_tech_delay}
}
\end{figure}

\begin{table}[h]
\centering
\scriptsize
\setlength{\tabcolsep}{0.15em}
\caption{Timing-aggressive 2D baselines, based on the \emph{OSU} libraries~\cite{OSU_PDK}.
All layouts are DRC clean. Area is in $\mu m^2$, power in $mW$, and delay in $ns$.
}
\smallerspacecaption
\begin{tabular}{|c|c|c|c|c|c|c|c|c|}
\hline
\multirow{2}{*}{\textbf{Benchmark}} & \multicolumn{4}{|c|}{\textbf{45nm}} & \multicolumn{4}{|c|}{\textbf{180nm}} \\
\cline{2-9}
 &
 \textbf{Instances} &
 \textbf{Area} &
 \textbf{Power} &
 \textbf{Delay} &
 \textbf{Instances} &
 \textbf{Area} &
 \textbf{Power} &
 \textbf{Delay} \\
 \hline \hline
b17\_1 &  
14,850 & 32,770.28  & 8.85  & 2.29  & 14,711  & 417,416 & 71.54 & 3.59 \\ \hline
b20 &  
6,959 & 15,549.31  & 8.12  & 2.87  & 7,521  & 216,168 & 97.94 & 3.6 \\ \hline
b21 &  
7,327 & 16,096.05  & 8.79  & 2.88 & 7,060  & 203,216 & 85.66 & 3.89 \\ \hline
\end{tabular}
\\ \
\\ \
The 45nm node is four generations away from 180nm, and delays improve by $\approx$30\% per generation~\cite{borkar99}; surprisingly, delay
degradations for \emph{OSU 180nm} are considerably off from this expectation.
We believe that this is due to the academic nature of the library.
\label{tab:basic_PPA_two_tech}
\end{table}


\begin{thebibliography}{10}
\providecommand{\url}[1]{#1}
\csname url@samestyle\endcsname
\providecommand{\newblock}{\relax}
\providecommand{\bibinfo}[2]{#2}
\providecommand{\BIBentrySTDinterwordspacing}{\spaceskip=0pt\relax}
\providecommand{\BIBentryALTinterwordstretchfactor}{4}
\providecommand{\BIBentryALTinterwordspacing}{\spaceskip=\fontdimen2\font plus
\BIBentryALTinterwordstretchfactor\fontdimen3\font minus
  \fontdimen4\font\relax}
\providecommand{\BIBforeignlanguage}[2]{{%
\expandafter\ifx\csname l@#1\endcsname\relax
\typeout{** WARNING: IEEEtran.bst: No hyphenation pattern has been}%
\typeout{** loaded for the language `#1'. Using the pattern for}%
\typeout{** the default language instead.}%
\else
\language=\csname l@#1\endcsname
\fi
#2}}
\providecommand{\BIBdecl}{\relax}
\BIBdecl

\bibitem{patnaik18_3D_ICCAD}
S.~Patnaik, M.~Ashraf, O.~Sinanoglu, and J.~Knechtel, ``Best of both worlds:
  Integration of split manufacturing and camouflaging into a security-driven
  {CAD} flow for {3D} {ICs},'' in \emph{Proc. Int. Conf. Comp.-Aided Des.},
  2018.

\bibitem{lipp18}
\BIBentryALTinterwordspacing
M.~Lipp \emph{et~al.}, ``Meltdown,'' \emph{ArXiv e-prints}, 2018. [Online].
  Available: \url{https://arxiv.org/abs/1801.01207}
\BIBentrySTDinterwordspacing

\bibitem{xiao16}
\BIBentryALTinterwordspacing
K.~Xiao, D.~Forte, Y.~Jin, R.~Karri, S.~Bhunia, and M.~Tehranipoor, ``Hardware
  trojans: Lessons learned after one decade of research,'' \emph{Trans. Des.
  Autom. Elec. Sys.}, vol.~22, no.~1, pp. 6:1--6:23, 2016. [Online]. Available:
  \url{http://doi.acm.org/10.1145/2906147}
\BIBentrySTDinterwordspacing

\bibitem{quadir16}
S.~E. Quadir \emph{et~al.}, ``A survey on chip to system reverse engineering,''
  \emph{J. Emerg. Tech. Comp. Sys.}, vol.~13, no.~1, pp. 6:1--6:34, 2016.

\bibitem{knechtel2019protect}
J.~Knechtel, S.~Patnaik, and O.~Sinanoglu, ``Protect your chip design
  intellectual property: An overview,'' \emph{Computing Research Repository},
  vol. abs/1902.05333, 2019.

\bibitem{knechtel17_TSLDM}
J.~Knechtel, O.~Sinanoglu, I.~A.~M. Elfadel, J.~Lienig, and C.~C.~N. Sze,
  ``Large-scale {3D} chips: Challenges and solutions for design automation,
  testing, and trustworthy integration,'' \emph{Trans. Sys. {LSI} Des.
  Method.}, vol.~10, pp. 45--62, 2017, invited paper.

\bibitem{fick13}
D.~Fick \emph{et~al.}, ``{Centip3De}: A cluster-based {NTC} architecture with
  64 {ARM Cortex-M3} cores in {3D} stacked 130 nm {CMOS},'' \emph{J. Sol.-St.
  Circ.}, vol.~48, no.~1, pp. 104--117, 2013.

\bibitem{kim12_3dmaps}
D.~H. Kim \emph{et~al.}, ``{3D-MAPS}: {3D} massively parallel processor with
  stacked memory,'' in \emph{Proc. Int. Sol.-St. Circ. Conf.}, 2012, pp.
  188--190.

\bibitem{jiang18}
\BIBentryALTinterwordspacing
H.~Jiang \emph{et~al.}, ``A provable key destruction scheme based on memristive
  crossbar arrays,'' \emph{Nature Electronics}, vol.~1, no.~10, pp. 548--554,
  2018. [Online]. Available: \url{https://doi.org/10.1038/s41928-018-0146-5}
\BIBentrySTDinterwordspacing

\bibitem{imeson13}
\BIBentryALTinterwordspacing
F.~Imeson, A.~Emtenan, S.~Garg, and M.~V. Tripunitara, ``Securing computer
  hardware using {3D} integrated circuit ({IC}) technology and split
  manufacturing for obfuscation,'' in \emph{Proc. USENIX Sec. Symp.}, 2013, pp.
  495--510. [Online]. Available:
  \url{\url{https://www.usenix.org/system/files/conference/usenixsecurity13/sec13-paper_imeson.pdf}}
\BIBentrySTDinterwordspacing

\bibitem{li18}
M.~Li, B.~Yu, Y.~Lin, X.~Xu, W.~Li, and D.~Z. Pan, ``A practical split
  manufacturing framework for trojan prevention via simultaneous wire lifting
  and cell insertion,'' in \emph{Proc. Asia South Pac. Des. Autom. Conf.},
  2018, pp. 265--270.

\bibitem{ku18}
B.~W. Ku, K.~Chang, and S.~K. Lim, ``{Compact-2D}: A physical design
  methodology to build commercial-quality face-to-face-bonded {3D} {ICs},'' in
  \emph{Proc. Int. Symp. Phys. Des.}, 2018, pp. 90--97.

\bibitem{chang16}
K.~Chang \emph{et~al.}, ``{Cascade2D}: A design-aware partitioning approach to
  monolithic {3D IC} with {2D} commercial tools,'' in \emph{Proc. Int. Conf.
  Comp.-Aided Des.}, 2016, pp. 130:1--130:8.

\bibitem{peng17}
Y.~Peng, T.~Song, D.~Petranovic, and S.~K. Lim, ``Parasitic extraction for
  heterogeneous face-to-face bonded {3-D} {ICs},'' \emph{Trans. Compon., Pack.,
  Manuf. Tech.}, vol.~7, no.~6, pp. 912--924, 2017.

\bibitem{rajendran13_split}
J.~Rajendran, O.~Sinanoglu, and R.~Karri, ``Is split manufacturing secure?'' in
  \emph{Proc. Des. Autom. Test Europe}, 2013, pp. 1259--1264.

\bibitem{wang18_SM}
Y.~Wang, P.~Chen, J.~Hu, G.~Li, and J.~Rajendran, ``The cat and mouse in split
  manufacturing,'' \emph{Trans. VLSI Syst.}, vol.~26, no.~5, pp. 805--817,
  2018.

\bibitem{patnaik18_SM_DAC}
S.~Patnaik, M.~Ashraf, J.~Knechtel, and O.~Sinanoglu, ``Raise your game for
  split manufacturing: Restoring the true functionality through {BEOL},'' in
  \emph{Proc. Des. Autom. Conf.}, 2018, pp. 140:1--140:6.

\bibitem{patnaik18_SM_ASPDAC}
S.~Patnaik, J.~Knechtel, M.~Ashraf, and O.~Sinanoglu, ``Concerted wire lifting:
  Enabling secure and cost-effective split manufacturing,'' in \emph{Proc. Asia
  South Pac. Des. Autom. Conf.}, 2018, pp. 251--258.

\bibitem{tezzaron08}
\BIBentryALTinterwordspacing
{Tezzaron Semiconductor}, ``{3D-ICs} and integrated circuit security,''
  Tezzaron Semiconductor, Tech. Rep., 2008. [Online]. Available:
  \url{http://tezzaron.com/media/3D-ICs_and_Integrated_Circuit_Security.pdf}
\BIBentrySTDinterwordspacing

\bibitem{dofe17}
J.~Dofe, P.~Gu, D.~Stow, Q.~Yu, E.~Kursun, and Y.~Xie, ``Security threats and
  countermeasures in three-dimensional integrated circuits,'' in \emph{Proc.
  Great Lakes Symp. VLSI}, 2017, pp. 321--326.

\bibitem{xie17}
Y.~Xie, C.~Bao, and A.~Srivastava, ``Security-aware {2.5D} integrated circuit
  design flow against hardware {IP} piracy,'' \emph{Computer}, vol.~50, no.~5,
  pp. 62--71, 2017.

\bibitem{gu2018cost}
P.~Gu, D.~Stow, P.~Mukim, S.~Li, and Y.~Xie, ``Cost-efficient {3D} integration
  to hinder reverse engineering during and after manufacturing,'' in
  \emph{Proc. Asian Hardw.-Orient. Sec. Trust Symp.}, 2018, pp. 74--79.

\bibitem{yan17_camo}
C.~Yan, J.~Dofe, S.~Kontak, Q.~Yu, and E.~Salman, ``Hardware-efficient logic
  camouflaging for monolithic {3D} {ICs},'' \emph{Trans. Circ. Sys.}, vol.~65,
  no.~6, pp. 799--803, 2018.

\bibitem{rajendran13_camouflage}
J.~Rajendran, M.~Sam, O.~Sinanoglu, and R.~Karri, ``Security analysis of
  integrated circuit camouflaging,'' in \emph{Proc. Comp. Comm. Sec.}, 2013,
  pp. 709--720.

\bibitem{wang16_MUX}
X.~Wang \emph{et~al.}, ``Secure and low-overhead circuit obfuscation technique
  with multiplexers,'' in \emph{Proc. Great Lakes Symp. VLSI}, 2016, pp.
  133--136.

\bibitem{nirmala16}
I.~R. Nirmala, D.~Vontela, S.~Ghosh, and A.~Iyengar, ``A novel threshold
  voltage defined switch for circuit camouflaging,'' in \emph{Proc. Europe
  Test. Symp.}, 2016, pp. 1--2.

\bibitem{akkaya18}
N.~E.~C. Akkaya, B.~Erbagci, and K.~Ma, ``A secure camouflaged logic family
  using postmanufacturing programming with a {3.6GHz} adder prototype in 65nm
  {CMOS} at {1V} nominal {VDD},'' in \emph{Proc. Int. Sol.-St. Circ. Conf.},
  2018.

\bibitem{patnaik17_Camo_BEOL_ICCAD}
S.~Patnaik, M.~Ashraf, J.~Knechtel, and O.~Sinanoglu, ``Obfuscating the
  interconnects: Low-cost and resilient full-chip layout camouflaging,'' in
  \emph{Proc. Int. Conf. Comp.-Aided Des.}, 2017, pp. 41--48.

\bibitem{subramanyan15}
P.~Subramanyan, S.~Ray, and S.~Malik, ``Evaluating the security of logic
  encryption algorithms,'' in \emph{Proc. Int. Symp. Hardw.-Orient. Sec.
  Trust}, 2015, pp. 137--143.

\bibitem{li16_camouflaging}
M.~Li \emph{et~al.}, ``Provably secure camouflaging strategy for {IC}
  protection,'' in \emph{Proc. Int. Conf. Comp.-Aided Des.}, 2016, pp.
  28:1--28:8.

\bibitem{yasin16_CamoPerturb}
M.~Yasin, B.~Mazumdar, O.~Sinanoglu, and J.~Rajendran, ``{CamoPerturb}: Secure
  {IC} camouflaging for minterm protection,'' in \emph{Proc. Int. Conf.
  Comp.-Aided Des.}, 2016, pp. 29:1--29:8.

\bibitem{guimaraes17}
L.~A. Guimar{\~a}es, R.~P. Bastos, and L.~Fesquet, ``Detection of layout-level
  trojans by monitoring substrate with preexisting built-in sensors,'' in
  \emph{Proc. Comp. Soc. Symp. VLSI}, 2017, pp. 290--295.

\bibitem{basak17}
A.~Basak, S.~Bhunia, T.~Tkacik, and S.~Ray, ``Security assurance for
  system-on-chip designs with untrusted {IPs},'' \emph{Trans. Inf. Forens.
  Sec.}, vol.~12, no.~7, pp. 1515--1528, 2017.

\bibitem{shi17_OBISA}
Q.~Shi, K.~Xiao, D.~Forte, and M.~M. Tehranipoor, ``Securing split manufactured
  {ICs} with wire lifting obfuscated built-in self-authentication,'' in
  \emph{Proc. Great Lakes Symp. VLSI}, 2017, pp. 339--344.

\bibitem{garg13}
S.~Garg and D.~Marculescu, ``Mitigating the impact of process variation on the
  performance of {3-D} integrated circuits,'' \emph{Trans. VLSI Syst.},
  vol.~21, no.~10, pp. 1903--1914, 2013.

\bibitem{CEP_github}
\BIBentryALTinterwordspacing
(2019) Assistant secretary of defense for research and engineering. Common
  Evaluation Platform. [Online]. Available: \url{https://github.com/mit-ll/CEP}
\BIBentrySTDinterwordspacing

\bibitem{nangate11}
\BIBentryALTinterwordspacing
(2011) {NanGate FreePDK45 Open Cell Library}. Nangate Inc. [Online]. Available:
  \url{http://www.nangate.com/?page_id=2325}
\BIBentrySTDinterwordspacing

\bibitem{synopsys32nm}
\BIBentryALTinterwordspacing
(2019) Synopsys 90nm generic libraries. {Synopsys}. [Online]. Available:
  \url{https://www.synopsys.com/community/university-program/teaching-resources.html}
\BIBentrySTDinterwordspacing

\bibitem{jung17}
M.~Jung, T.~Song, Y.~Peng, and S.~K. Lim, ``Design methodologies for low-power
  {3-D} {ICs} with advanced tier partitioning,'' \emph{Trans. VLSI Syst.},
  vol.~25, no.~7, 2017.

\bibitem{chen15}
S.~Chen, J.~Chen, D.~Forte, J.~Di, M.~Tehranipoor, and L.~Wang, ``Chip-level
  anti-reverse engineering using transformable interconnects,'' in \emph{Proc.
  Int. Symp. Def. Fault Tol. in VLSI Nanotech. Sys.}, 2015, pp. 109--114.

\bibitem{uhlig18}
B.~Uhlig \emph{et~al.}, ``Progress on carbon nanotube {BEOL} interconnects,''
  in \emph{Proc. Des. Autom. Test Europe}, 2018.

\bibitem{webinterface}
\BIBentryALTinterwordspacing
(2017) {3D SM Attack by DfX Lab, NYUAD}. [Online]. Available:
  \url{https://github.com/DfX-NYUAD/3D-SM-Attack}
\BIBentrySTDinterwordspacing

\bibitem{Opencores}
\BIBentryALTinterwordspacing
(2019) {R}eference community for {F}ree and {O}pen {S}ource {IP} cores.
  Opencores. [Online]. Available: \url{https://opencores.org/}
\BIBentrySTDinterwordspacing

\bibitem{salmani16}
H.~Salmani and M.~M. Tehranipoor, ``Vulnerability analysis of a circuit layout
  to hardware trojan insertion,'' \emph{Trans. Inf. Forens. Sec.}, vol.~11,
  no.~6, pp. 1214--1225, 2016.

\bibitem{cheng10}
J.~Cheng, A.~W.-c. Fu, and J.~Liu, ``K-isomorphism: Privacy preserving network
  publication against structural attacks,'' in \emph{Proc. SIGMOD}, 2010, pp.
  459--470.

\end{thebibliography}

\begin{thebibliography}{10}
\providecommand{\url}[1]{#1}
\csname url@samestyle\endcsname
\providecommand{\newblock}{\relax}
\providecommand{\bibinfo}[2]{#2}
\providecommand{\BIBentrySTDinterwordspacing}{\spaceskip=0pt\relax}
\providecommand{\BIBentryALTinterwordstretchfactor}{4}
\providecommand{\BIBentryALTinterwordspacing}{\spaceskip=\fontdimen2\font plus
\BIBentryALTinterwordstretchfactor\fontdimen3\font minus
  \fontdimen4\font\relax}
\providecommand{\BIBforeignlanguage}[2]{{%
\expandafter\ifx\csname l@#1\endcsname\relax
\typeout{** WARNING: IEEEtran.bst: No hyphenation pattern has been}%
\typeout{** loaded for the language `#1'. Using the pattern for}%
\typeout{** the default language instead.}%
\else
\language=\csname l@#1\endcsname
\fi
#2}}
\providecommand{\BIBdecl}{\relax}
\BIBdecl

\bibitem{OSU_PDK}
\BIBentryALTinterwordspacing
(2017) {FreePDK: Unleashing VLSI to the Masses}. Oklahoma State University.
  [Online]. Available: \url{https://vlsiarch.ecen.okstate.edu/flows/}
\BIBentrySTDinterwordspacing

\bibitem{borkar99}
S.~Borkar, ``Design challenges of technology scaling,'' \emph{IEEE Micro},
  vol.~19, no.~4, pp. 23--29, 1999.

\end{thebibliography}
\end{document}